\documentclass[10pt,sigconf]{acmart}
\usepackage[T1]{fontenc}
\usepackage{tabularx}
\usepackage{subfigure}
\usepackage{wrapfig}
\usepackage{pbox}
\usepackage{bbding}
\usepackage{textcomp}
\usepackage{xcolor}
\usepackage[noend]{algorithm2e}
\usepackage{bbm}
\usepackage{dblfloatfix}
\usepackage{pifont}
\usepackage{bm}
\usepackage{enumitem}
\usepackage{duckuments}

\usepackage{setspace}
\usepackage{amsmath}
\usepackage{lipsum}

\ifodd 0
\definecolor{darkgreen}{RGB}{0,200,0}
\newcommand{\rev}[1]{{\color{blue}#1}} 
\newcommand{\newrev}[1]{{\color{red}#1}} 
\else
\newcommand{\rev}[1]{#1}
\newcommand{\newrev}[1]{#1} 

\fi

\def\BibTeX{{\rm B\kern-.05em{\sc i\kern-.025em b}\kern-.08em
    T\kern-.1667em\lower.7ex\hbox{E}\kern-.125emX}}

\setcopyright{acmcopyright}
\acmYear{2025}
\copyrightyear{2025}
\setcopyright{rightsretained}
\acmConference[ACM MM'25]{The 33rd ACM International Conference on Multimedia}{October, 2025}{Dublin, Ireland}
\acmBooktitle{The 33rd ACM International Conference on Multimedia (ACM MM'25), October, 2025, Dublin, Ireland}
\acmDOI{10.1145/3570361.3613290}
\acmISBN{978-1-4503-9990-6/23/10}

\usepackage{glossaries}
\usepackage[symbols,nogroupskip]{glossaries-extra}

\newglossaryentry{tar}
{
	name={\ensuremath{\mathcal S}},
	sort={p},
	description={Subscript of Target User}
}

\newglossaryentry{int}
{
	name={\ensuremath{\mathcal I}},
	sort={p},
	description={Subscript of Interfering User}
}

\newglossaryentry{int_i}
{
	name={\ensuremath{\mathcal I_i}},
	sort={p},
	description={Subscript of Interfering User}
}

\newglossaryentry{int_j}
{
	name={\ensuremath{\mathcal I_j}},
	sort={p},
	description={Subscript of Interfering User}
}

\newglossaryentry{num_ue_ant}
{
	name={\ensuremath{A_{\mathrm{ue}}}},
	sort={p},
	description={Number of UE's antenna}
}

\newglossaryentry{num_ap_ant}
{
	name={\ensuremath{A_{\mathrm{ap}}}},
	sort={p},
	description={Number of AP's antenna}
}

\newglossaryentry{numAddr}
{
	name={\ensuremath{N_{\mathrm{mac}}}},
	sort={p},
	description={Number of AP's antenna}
}

\newglossaryentry{numInterferer}
{
	name={\ensuremath{N_{\gls{int}}}},
	sort={p},
	description={Number of AP's antenna}
}

\newglossaryentry{setInterferer}
{
	name={\ensuremath{\{1,... N-1\}}},
	sort={p},
	description={Number of AP's antenna}
}

\newglossaryentry{threshold}
{
	name={\ensuremath{\gamma_{\mathrm{th}}}},
	sort={p},
	description={Number of AP's antenna}
}

\newglossaryentry{numSubCarrier}
{
	name={\ensuremath{N_{\mathrm{SC}}}},
	sort={p},
	description={Number of subcarriers}
}

\newglossaryentry{numSTS}
{
	name={\ensuremath{N_{\mathrm{STS}}}},
	sort={p},
	description={Number of space time stream}
}

\newglossaryentry{numPhase}
{
	name={\ensuremath{N_{\mathrm{P}}}},
	sort={p},
	description={Number of Rx antennas}
}

\newglossaryentry{numFragments}
{
	name={\ensuremath{N_{\mathrm{F}}}},
	sort={p},
	description={Number of Rx antennas}
}

\newglossaryentry{rsampleFreq}
{
	name={\ensuremath{f_{\mathrm{rs}}}},
	sort={p},
	description={Frequency of resampling}
}

\newglossaryentry{cutoffFreq}
{
	name={\ensuremath{f_{\mathrm{cut}}}},
	sort={p},
	description={Frequency of resampling}
}

\newglossaryentry{nspThreshold_N}
{
	name={\ensuremath{N_{\mathrm{nsp}}}},
	sort={p},
	description={Frequency of resampling}
}

\newglossaryentry{nspThreshold_t}
{
	name={\ensuremath{\Delta t}},
	sort={p},
	description={Frequency of resampling}
}

\newglossaryentry{nspThreshold_T}
{
	name={\ensuremath{T_{\mathrm{t}}}},
	sort={p},
	description={Frequency of resampling}
}

\newglossaryentry{num_nsp_slice}
{
	name={\ensuremath{F_{\mathrm{nsp}}}},
	sort={p},
	description={Number of non-sparse time slices}
}

\newglossaryentry{num_sp_slice}
{
	name={\ensuremath{F_{\mathrm{sp}}}},
	sort={p},
	description={Number of sparse time slices}
}

\newglossaryentry{freq_resample}
{
	name={\ensuremath{f_{\mathrm{rs}}}},
	sort={p},
	description={Number of sparse time slices}
}

\newglossaryentry{numHid}
{
	name={\ensuremath{H}},
	sort={p},
	description={Number of hidden layers}
}

\newglossaryentry{filterDim}
{
	name={\ensuremath{L_{\mathrm{kn}}}},
	sort={p},
	description={size of kernel filter}
}

\newglossaryentry{numResBlock}
{
	name={\ensuremath{L_{\mathrm{f}}}},
	sort={p},
	description={size of kernel filter}
}

\newglossaryentry{numResuse}
{
	name={\ensuremath{N_{\mathrm{re}}}},
	sort={p},
	description={size of kernel filter}
}

\newglossaryentry{numChannel}
{
	name={\ensuremath{N_{\mathrm{ch}}}},
	sort={p},
	description={size of channel}
}

\newglossaryentry{numFreqCp}
{
	name={\ensuremath{N_{\mathrm{F}}}},
	sort={p},
	description={number of frequency components}
}

\newglossaryentry{numSample}
{
	name={\ensuremath{N_{\mathrm{s}}}},
	sort={p},
	description={size of channel}
}

\newglossaryentry{numTx}
{
	name={\ensuremath{N_{\mathrm{tx}}}},
	sort={p},
	description={Number of UE's antenna}
}
\newglossaryentry{numRx}
{
	name={\ensuremath{N_{\mathrm{rx}}}},
	sort={p},
	description={Number of UE's antenna}
}

\begin{document}

\newcommand{\name}{S\textsc{pace}V\textsc{erse}\xspace}
\title{A Satellite-Ground Synergistic Large Vision-Language Model System for \newrev{Earth Observation}}

 
 

\author{ {\Large
    Yuxin Zhang$^{1,2}$
    \quad Jiahao Yang$^{1,2}$ \quad Zhe Chen$^{1}$ \quad 
 Wenjun Zhu$^{1,2}$ \quad Jin Zhao$^{1,2}$\quad Yue Gao$^{1}$
    }}
    \thanks{Accepted at ACM Multimedia 2025.}
\affiliation{
   \institution{{\normalsize
	$^1$Institute of Space Internet, Fudan University, Shanghai \country{China} \\
	$^2$College of Computer Science and Artificial Intelligence, Fudan University, Shanghai \country{China} \\
    Email: \{yxzhang24, 24210240374\}@m.fudan.edu.cn, \{zhechen, wenjun, jzhao, gao.yue\}@fudan.edu.cn
        }}
    }
    
    \renewcommand{\authors}{Y. Zhang, J. Yang, Z. Chen, W, Zhu, J Zhao, and Y. Gao}
    \renewcommand{\shortauthors}{Y. Zhang, J. Yang, Z. Chen, W, Zhu, J Zhao, and Y. Gao}
\begin{abstract}
\newrev{Recently, large vision-language models~(LVLMs) unleash powerful analysis capabilities for low Earth orbit~(LEO) satellite Earth observation images in the data center. However, fast satellite motion, brief satellite-ground station~(GS) contact windows, and large size of the images pose a data download challenge. To enable near real-time Earth observation applications~(e.g., \rev{disaster and extreme weather monitoring}), we should explore how to deploy LVLM in LEO satellite networks, and design \name, an efficient satellite-ground synergistic LVLM inference system. To this end, firstly, we deploy compact LVLMs on satellites for lightweight tasks, whereas regular LVLMs operate on GSs to handle computationally intensive tasks. Then, we propose a computing and communication co-design framework comprised of a progressive confidence network, and an attention-based multi-scale preprocessing, used to identify on-satellite inferring data, and reduce data redundancy before satellite-GS transmission, separately.
}
\rev{
We implement, and evaluate \name on real-world \newrev{LEO satellite} constellations and datasets, achieving a 31.2\% average gain in accuracy and a 51.2\% reduction in latency compared to state-of-the-art baselines.
}
\end{abstract}



\maketitle

\section{Introduction}\label{sec:introduction} 

Advances in satellite technology have significantly lowered the cost of launches, leading to a surge in the deployment of low Earth orbit~(LEO) satellites~\cite{liu2024democratizing,fang2024robust,singh2024spectrumize,zhang2024satfed,lin2025leo,lin2025esl}.
\rev{For instance, SpaceX and Planet Labs are establishing mega-constellations by deploying hundreds to thousands of LEO satellites into orbit~\cite{ahmmed2022digital,planet_lab,lin2025leo,yuan2024satsense,peng2025sigchord,lin2024fedsn,zhao2024leo}~{(< 1000\!~km above Earth)~\cite{li2023networking}}.}
Equipped with advanced multimodal sensors (e.g., visual, and infrared), those satellites capitalize on their unique vantage point to capture high-resolution \newrev{Earth observation (a.k.a. remote sensing)} imagery 
from the Earth's surface~\cite{singh2024spectrumize,shenoy2024s4,hu2024utilizing,lin2025esl}.
\newrev{Through deep learning~(DL)-based image analysis, satellite data can effectively support a wide variety of space-based applications, such as}
weather forecasting~\cite{dewitte2021artificial,bhaga2020impacts}, transportation management~\cite{yang2020basic,zheng2023simultaneous}, and disaster monitoring~\cite{franch2020spatial,zhao2023seeing}




\rev{However, most of existing DL-based approaches for Earth observation tasks \newrev{depend heavily on convolutional neural networks (CNNs)
}
}~\cite{rolf2021generalizable,li2023large,lin2024efficient,fang2024ic3m,chen2021rf,peng2024sums,yuan2025constructing,hu2024accelerating,tang2024merit}, which are increasingly revealing efficiency and performance bottlenecks.
On one hand, \newrev{task-specific CNN model design and retraining require substantial human effort~\cite{sun2021convolutional,yuan2023graph,lin2025hasfl,wu2024netllm,zhang2025lcfed,chen2024gradient,lin2024adaptsfl}, hindering generalization across diverse tasks. On the other hand, single-modality CNN models with limited parameters \rev{(typically tens of millions)} constrain their representation capacity~\cite{kaplan2020scaling, 10490262,lin2025hsplitlora}.}
\newrev{To overcome the above CNN limitations, recent studies~\cite{liu2024remoteclip,zhang2024earthgpt,zhan2024skyeyegpt} have introduced multimodal \textit{large vision-language models} (LVLMs)~\cite{zhang2024vision,zhang2024pip,wang2024break,fang2024automated,lin2024splitlora} with over one billion parameters as a "one-model-for-all-task" framework~\cite{ramesh2021zero} that uses natural language prompts to perform Earth observation tasks (e.g., object detection and tracking)~\cite{kuckreja2024geochat}.
}

\begin{figure}[t]
\centering
\includegraphics[width=.96\columnwidth]{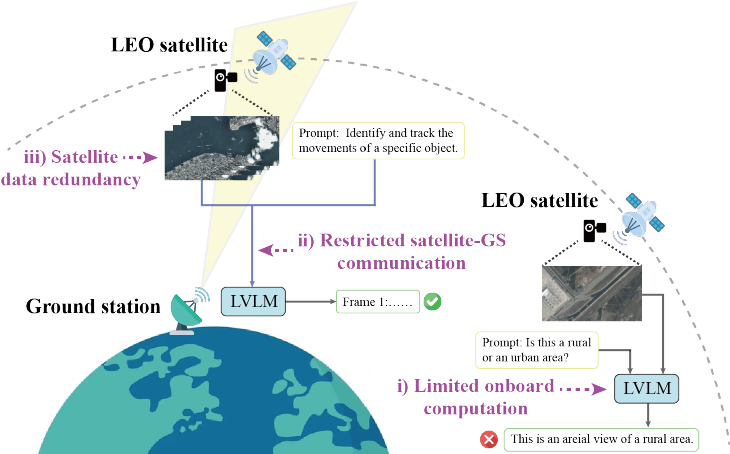}
\caption{
While LVLMs hold potential for Earth observation, their efficient deployment in satellite networks faces challenges including i) limited onboard computational capacity, ii) restricted satellite-GS connectivity, and iii) the high redundancy inherent in satellite data.}
\label{fig:vlm_over_satellite_network}
\vspace{-1em}
\end{figure}

\textbf{Challenges.}
\rev{Despite the promising of LVLMs, their near real-time deployment for Earth observation remains underexplored, presenting three key challenges as shown in Figure~\ref{fig:vlm_over_satellite_network}.}
\textit{Firstly}, satellites' strict computational constraints in power, size, and weight~\cite{denby2020orbital,raspberrypi} make deploying \rev{regular} LVLMs (e.g., Qwen2-VL-7B~\cite{wang2024qwen2}) infeasible, allowing only compact models (<3B parameters) designed for power-limited edge devices, which severely degrades \rev{accuracy}.
\textit{Secondly}, although abundant ground resources allow the deployment of larger, more powerful LVLMs, constrained satellite-ground station (GS) connectivity incurs substantial transmission latency~\cite{zhang2025s,vasisht2021l2d2,lin2024fedsn}.
\textit{Lastly}, satellite data exhibit substantial inherent redundancy, as only small regions within images are typically relevant to specific tasks~\cite{liu2024remoteclip,xia2018dota}. \rev{As a result, transmitting entire satellite images leads to substantial bandwidth inefficiency.}
In \S~\ref{sec:background_motivation}, we present motivating experiments to thoroughly explore these challenges.

\textbf{Our solutions.}
In this paper,
we propose \name as a satellite-ground \textit{synergistic} LVLM inference framework for \rev{Earth observation}.
To tackle the dual challenges of limited satellite computation and substantial satellite-GS transmission latency, \name deploys a compact LVLM onboard to deliver rapid responses for simple samples, while a larger model on the ground enhances performance for more complex samples.
We construct a progressive confidence network that optimizes satellite-GS task allocation while minimizing onboard \rev{computational costs.
In addition, to mitigate redundancy in satellite data, we introduce an attention-guided multi-scale preprocessing mechanism that significantly reduces satellite-to-ground transmission volume while preserving inference accuracy.}
Our main contributions are summarized as follows:
\begin{itemize}
  \item To the best of our knowledge, \name is the first satellite-ground synergistic LVLM inference framework tailored for LEO satellite networks.
  \item We design a progressive onboard confidence network to achieve accurate satellite-GS task allocation with minimal satellite computational costs.
  \item
  \rev{We propose a multi-scale preprocessing guided by text-image attention to reduce transmission redundancy while preserving the ground inference accuracy.}
  \item We implement the \name prototype and conduct extensive experiments, demonstrating its superiority over state-of-the-art methods.
\end{itemize}

The rest of the paper is organized as follows.
\S~\ref{sec:background_motivation} motivates \name by revealing the design challenges.
\S~\ref{sec:design_ovewview} presents the system design.
\S~\ref{sec:eval} introduces the implementation, followed by performance evaluation.
Related works and technical limitations are discussed in \S~\ref{sec:related_work}.
Finally, conclusions are presented in \S~\ref{sec:conclusion}.


\section{Background and Motivation}  \label{sec:background_motivation}

We study key limitations in deploying LVLMs for \rev{Earth observation}, thus motivating the design of \name.




\begin{figure}[t]
    \vspace{-1em}
    \setlength\abovecaptionskip{6pt}
     \setlength\subfigcapskip{0pt}
     	\subfigure[Relative performance.]{	
        \centering
		\label{subfig:relative_performance}
		\includegraphics[width=.479\columnwidth]{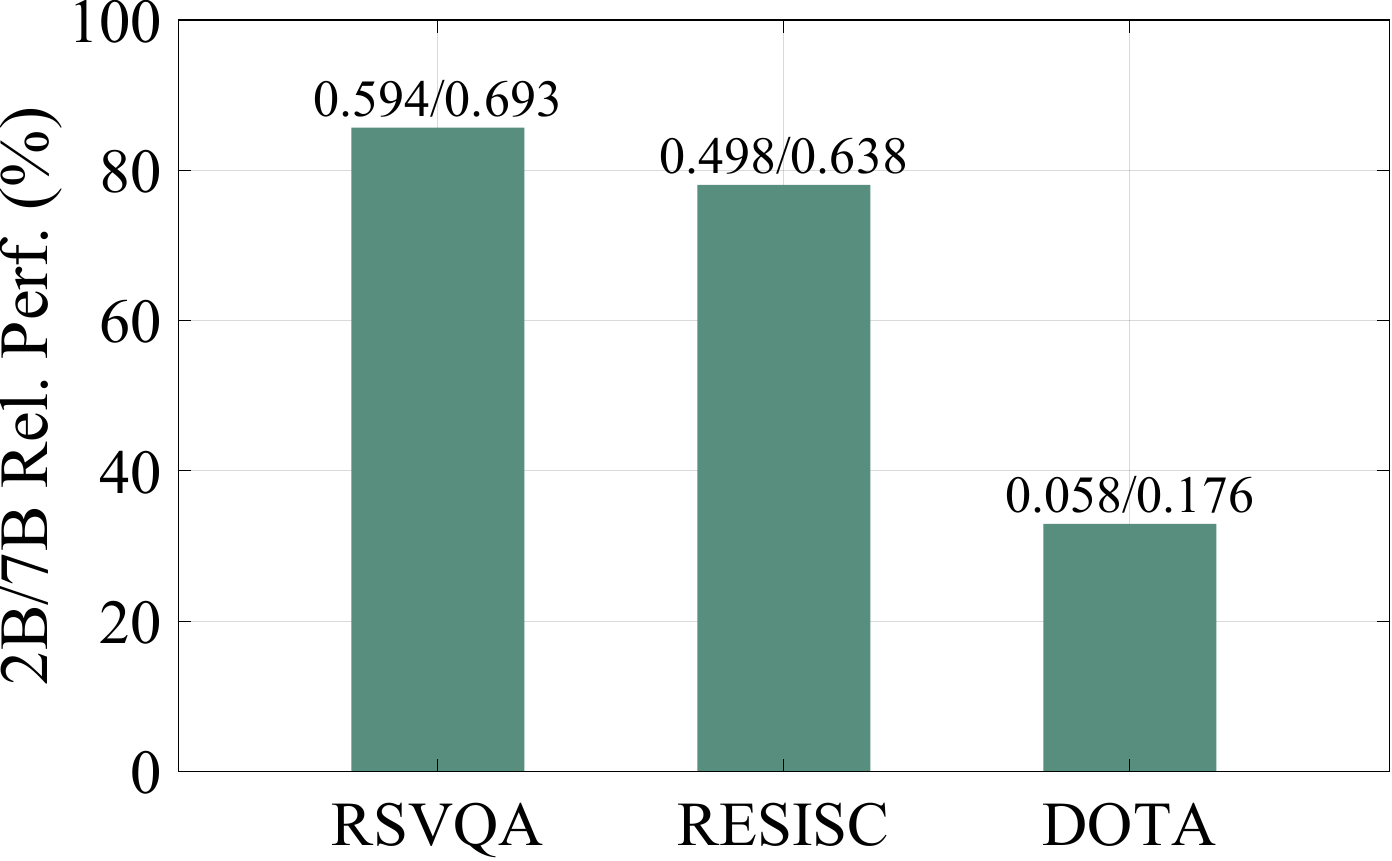}}
	\subfigure[Memory consumption.]{	
        \centering
		\label{subfig:memory_consumption}
		\includegraphics[width=.475\columnwidth]{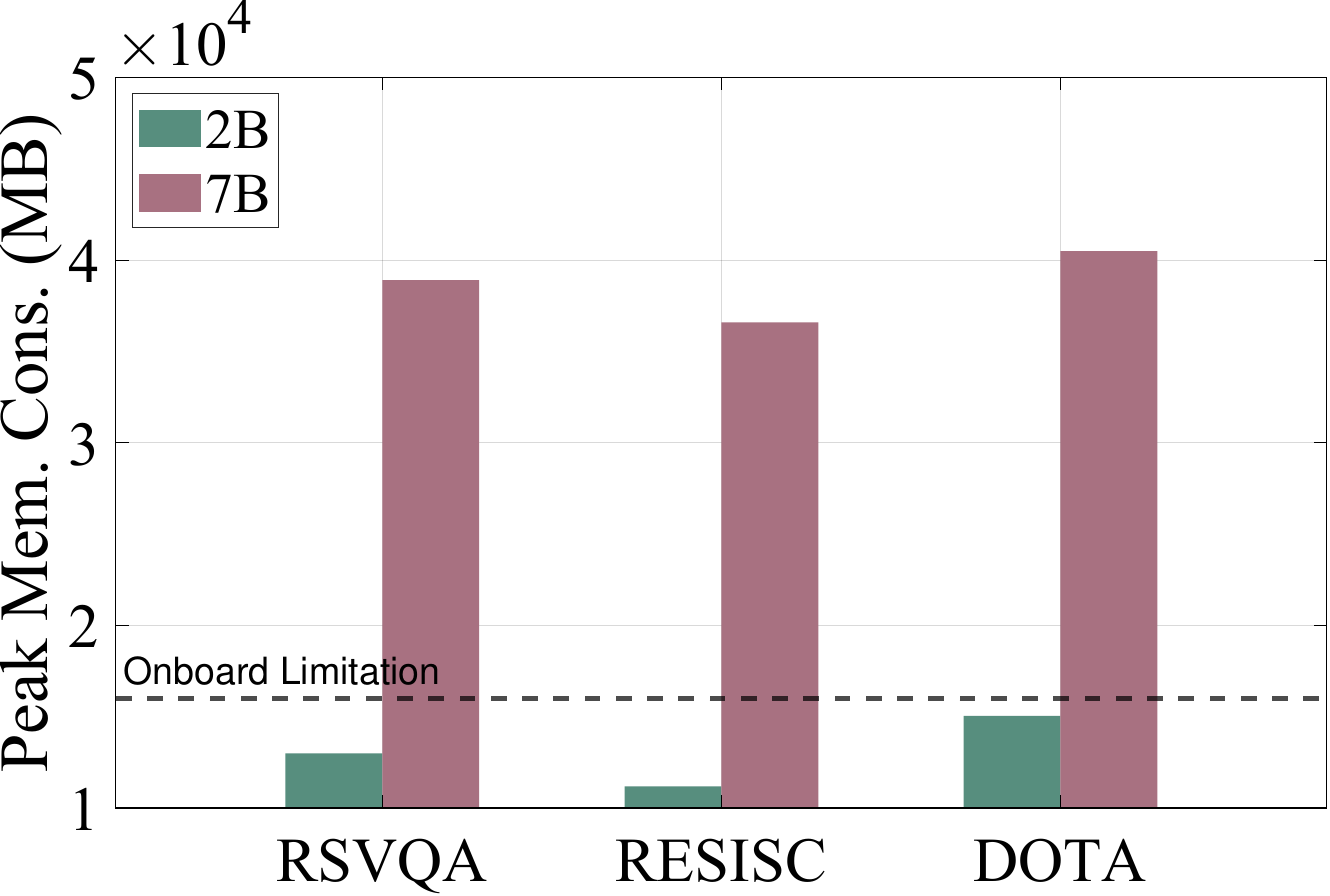}
	}
\caption{
The comparison of performance and memory consumption between the Qwen2-VL-2B and 7B models~\cite{wang2024qwen2} on emulated satellite platforms (with computation executed on 16 GB NVIDIA Jetson AGX Xavier).
}
    \label{fig:moti_onboard_model_size}
\vspace{-1em}
\end{figure}


\subsection{Limitations in Onboard Computation and Ground Connectivity}\label{subsec:limit_onboard_comp}

\subsubsection{Constrained Onboard Computation}
With the integration of commercial off-the-shelf computing devices in LEO satellites, onboard inference of DL models is increasingly viable~\cite{murphy2024deploying,manning2018machine}.
However, stringent constraints on size, weight, and power limit even flagship onboard systems~\cite{george2018onboard}, rendering them incapable of handling the computational demands of standard-sized LVLM workloads~\cite{10769058}.
We conduct experiments to assess onboard deployment of LVLMs with varying parameter scales, as shown in Figure~\ref{fig:moti_onboard_model_size}.
We deploy the widely recognized Qwen2-VL-7B~\cite{wang2024qwen2} LVLM (7 billion parameters), along with its compact counterpart, Qwen2-VL-2B (2 billion parameters), on emulated satellite platforms to evaluate their performance in three Earth observation tasks: \textit{visual question answering}, \textit{image classification}, and \textit{object detection}, leveraging the RSVQA LR~\cite{lobry2020rsvqa}, RESISC45~\cite{cheng2017remote}, and DOTA-v1.0~\cite{xia2018dota}, respectively.
We utilize a Kernel-based Virtual Machine (KVM) to emulate LEO satellites, with computation executed on separate 16GB NVIDIA Jetson AGX Xavier units, whose computational capacity closely resembles that of LEO satellite systems~\cite{jetson,george2018onboard}.

As shown in Figure~\ref{subfig:relative_performance}, the Qwen2-VL-7B model significantly outperforms its compact 2B counterpart, achieving an average performance gain of {82.7\%}.
However, as illustrated in Figure~\ref{subfig:memory_consumption}, the larger scale of the 7B model significantly increases deployment overhead, consuming an average of {24.9 GB} more memory than the 2B model and exceeding the satellite’s hardware limitations.
Thus, deploying a compact LVLM onboard is a necessary compromise, albeit at the cost of suboptimal performance.

\subsubsection{Intermittent Ground Connectivity}

\begin{figure}[t] 
    \vspace{-1em}
    \setlength\abovecaptionskip{6pt}
     \setlength\subfigcapskip{0pt}
     	\subfigure[Contact time.]{	
        \centering
		\label{subfig:motivating_contact}
		\includegraphics[width=.475\columnwidth]{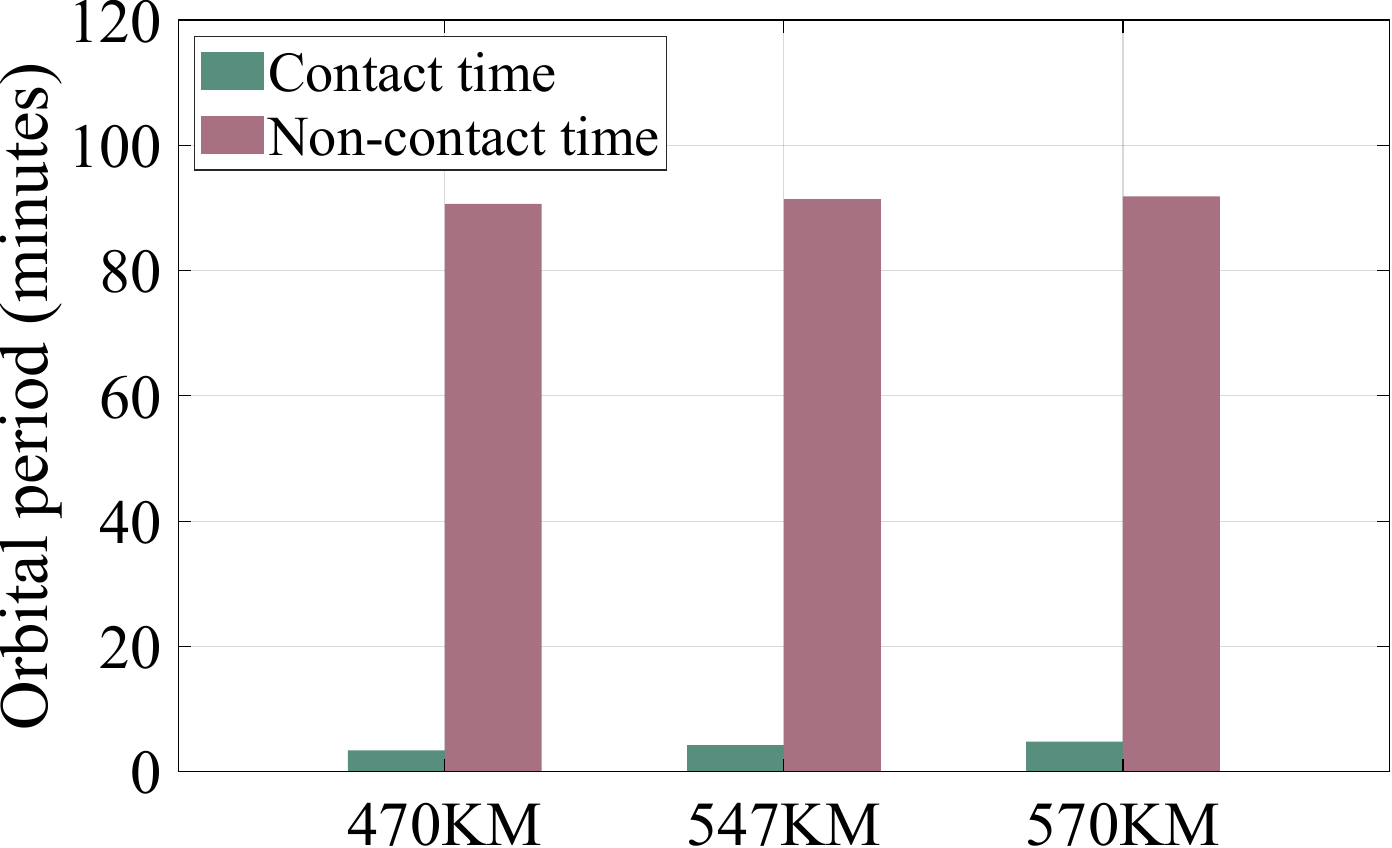}}
	\subfigure[Inference latency.]{	
        \centering
		\label{subfig:inference_delay}
		\includegraphics[width=.475\columnwidth]{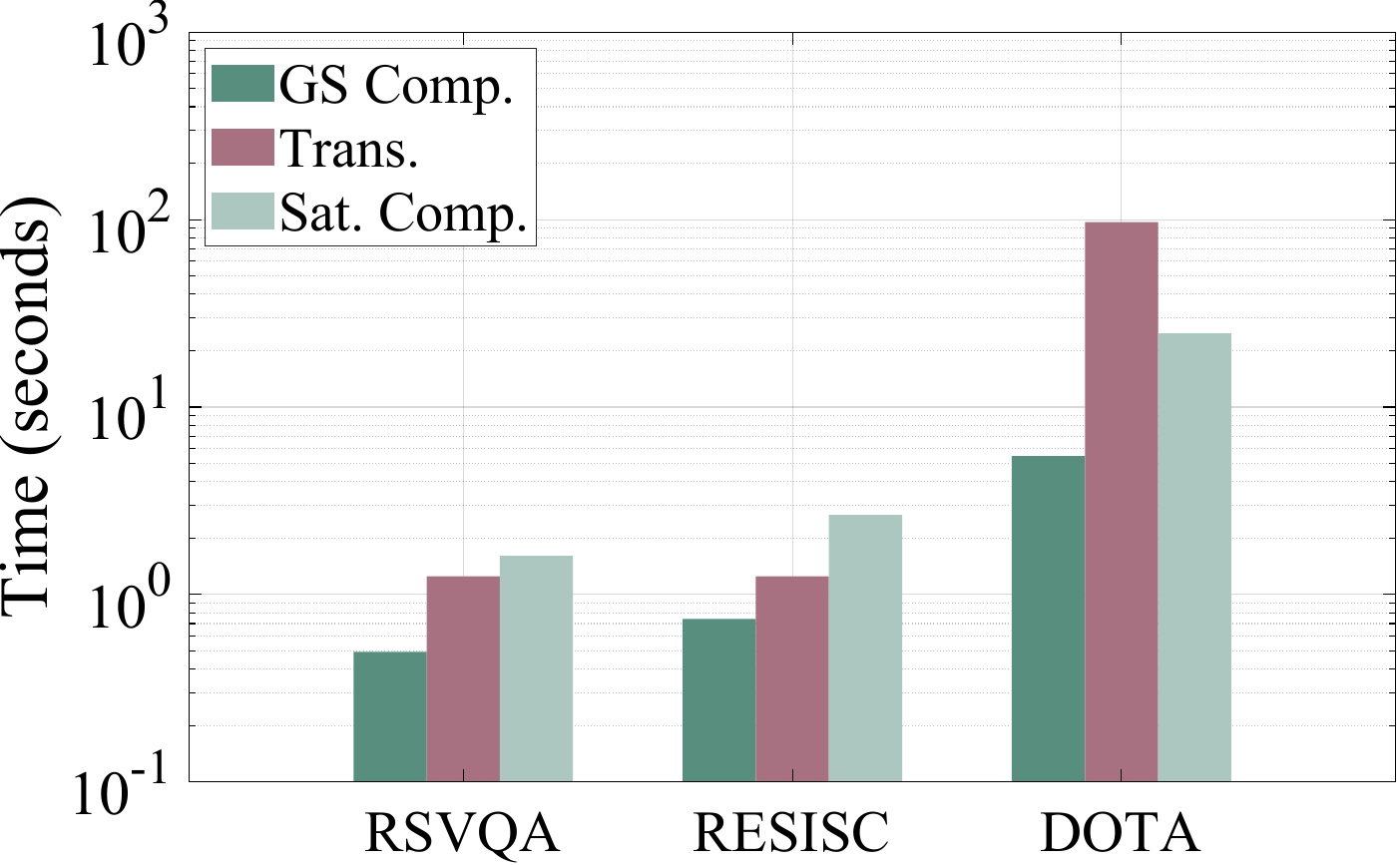}
	}
\caption{
Intermittent satellite-GS connectivity (a) causes significant delays when transmitting raw satellite data for GS inference (b).
}
    \label{fig:moti_ground_infer}
\vspace{-1em}
\end{figure}

We further investigate deploying LVLMs on the ground, where unrestricted computational resources allow for parameter-unconstrained models. However, efficiency is severely impacted by intermittent satellite-GS connectivity~\cite{10.1145/3570361.3592521}.
Based on the configuration data of the Starlink constellation~\cite{starlink_gp}, the satellite-GS contact duration can be derived at different orbital altitudes.
As shown in Figure~\ref{subfig:motivating_contact}, the contact window accounts for only an average 4.33\% of the orbital period.
We also deploy a Starlink GS (as described in \S~\ref{subsec:implement}) and measure downlink rates, observing an average of 110.67 Mbps.
Figure~\ref{subfig:inference_delay} presents experimental results on transmitting raw RS data to GS via satellite downlinks for Qwen2-VL-7B model inference.
Evidently, while ground inference delivers superior accuracy, data transmission accounts for a substantial 76.39\% of the total time, resulting in up to 4.14$\times$ the latency of onboard inference with the compact model—rendering it unacceptable for \rev{near real-time Earth observation applications.}

To conclude, the computational and communication constraints of real-world LEO satellite networks present pose significant challenges to efficient LVLM inference deployment:
i) \textit{onboard deployment} suffers from suboptimal performance due to resource limitations, while  
ii) \textit{ground deployment} incurs substantial inference latency due to intermittent connectivity.
These challenges highlight the need for a novel LVLM deployment paradigm tailored to LEO satellite networks, one that optimally balances performance and latency.

\begin{figure}[t] 
    \vspace{-1em}
    \setlength\abovecaptionskip{6pt}
     \setlength\subfigcapskip{0pt}
        \subfigure[Redundancy reduction.]{	
        \centering
		\label{subfig:random_mask}
		\includegraphics[width=.475\columnwidth]{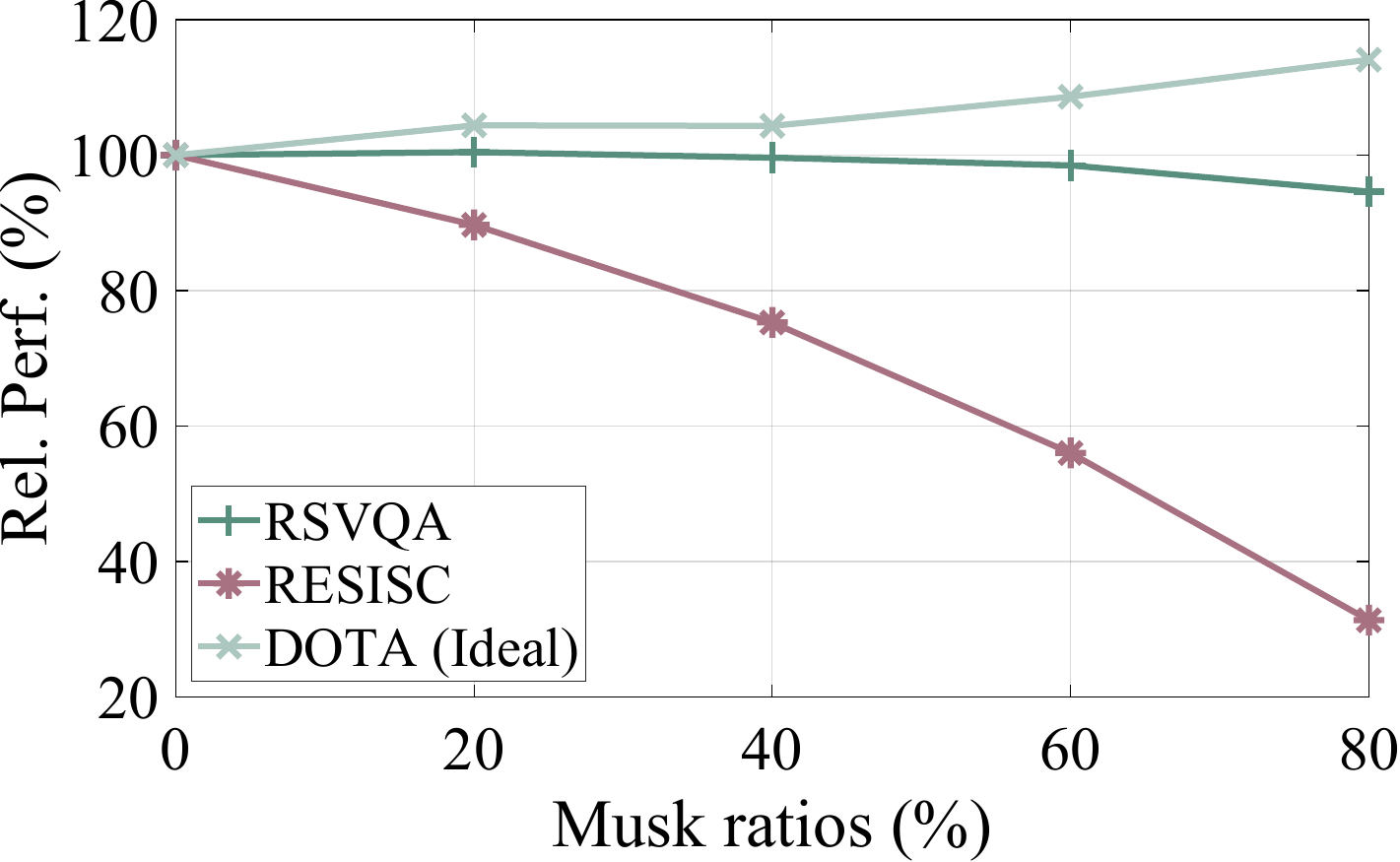}}
	\subfigure[Musk strategy on DOTA.]{
        \centering
		\label{subfig:musk_strategy}
		\includegraphics[width=.475\columnwidth]{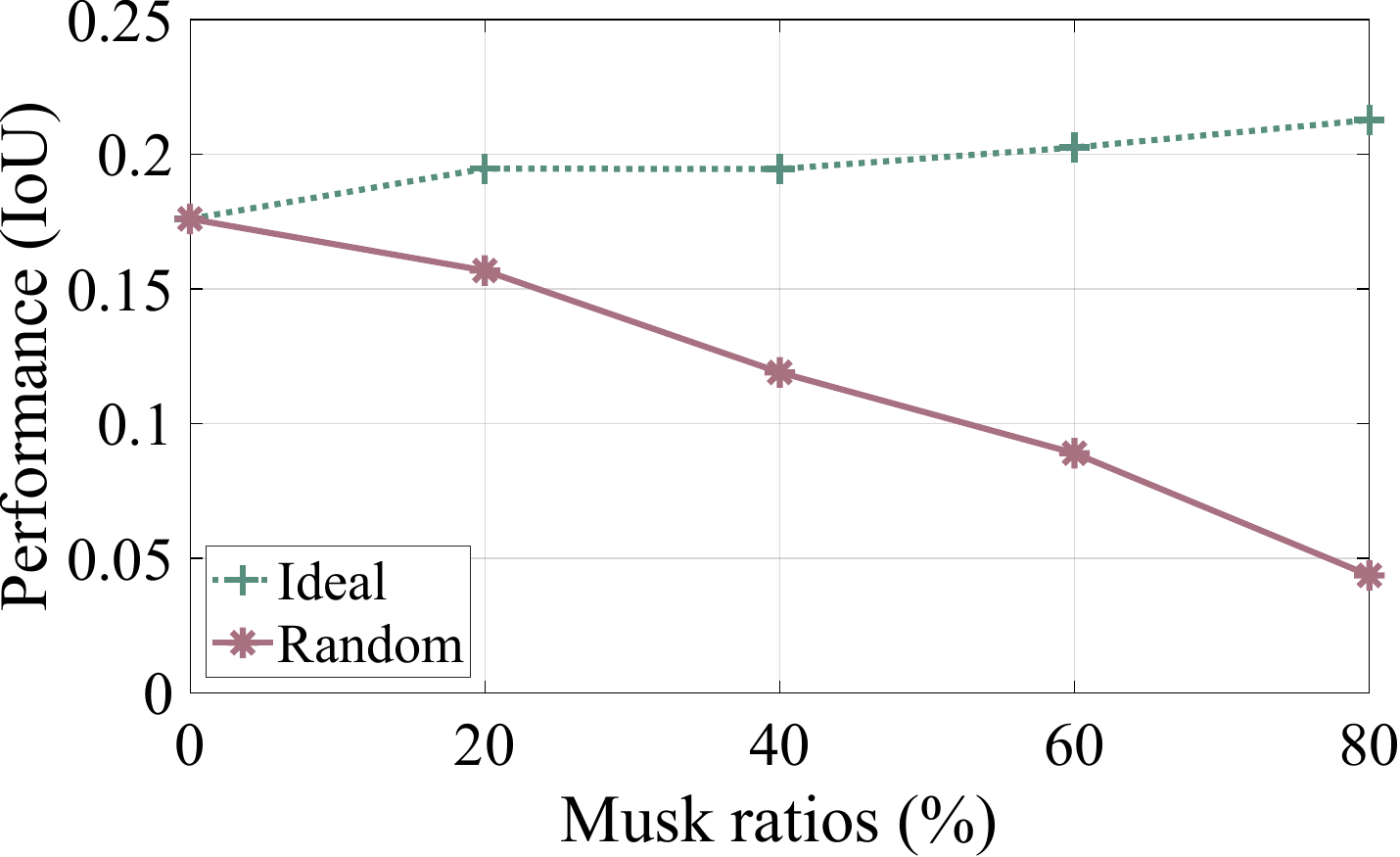}}
\caption{
\rev{Satellite images contain substantial redundancy (a); however, naive redundancy reduction strategies remain far from optimal (b)}.
}
    \label{fig:motivating_rs_redun}
\vspace{-1em}
\end{figure}

\subsection{High Redundancy of \rev{Satellite} Data}\label{subsec:rs_redundancy}

Beyond the inherent limitations of LEO satellite networks, substantial redundancy in satellite data further amplifies inefficiencies.
Specifically, \rev{Earth observation images} contain vast regions that are irrelevant to the given task (regions of non-interest), whose exclusion has minimal impact on inference results.
Consequently, transmitting full redundant data over satellite-GS links significantly increases inference latency and resource consumption without yielding any performance gains.

\rev{To better understand the inherent redundancy in satellite data, we conduct experiments by randomly masking at varying ratios during inference.
As shown in Figure~\ref{subfig:random_mask}, masking an average 40\% of satellite data results in only 6.92\% accuracy degradation.
}
Notably, in the object detection task, masking 80\% of the image—excluding target objects (an \textit{ideal} setting)—leads to a counterintuitive {14.09\%} accuracy (IoU) improvement, highlighting substantial internal redundancy in satellite data that may even interfere with other critical regions.
However, despite the vast presence of regions of non-interest, \rev{accurately} identifying them remains a fundamental challenge.
As shown in Figure~\ref{subfig:musk_strategy}, naive image redundancy reduction strategies, such as random masking, are suboptimal. On the DOTA dataset, an ideal masking strategy that selectively preserves target objects significantly outperforms random masking, which risks obscuring critical regions, resulting in severe performance degradation and reduced robustness.
Thus, it is essential to develop methods for accurately identifying and filtering regions of non-interest in satellite data, \rev{thereby reducing transmission volume in satellite networks} while preserving inference performance.

\begin{figure*}[t]
\centering
\includegraphics[width=.95\linewidth]{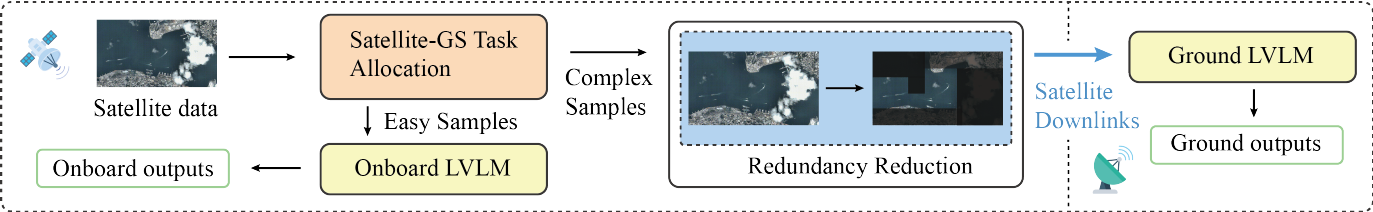}
\vspace{-.5em}
\caption{The \name framework overview consists of two key \rev{successive} components: i) allocating LVLM inference tasks between satellites and GS, and ii) \rev{mitigating satellite} data redundancy before transmission.
}
\label{fig:skyverse_frame}
\vspace{-1em}
\end{figure*}


\section{System Design}
\label{sec:design_ovewview}

In this section, we present \name, a satellite-ground synergistic LVLM inference system, as illustrated in Figure~\ref{fig:skyverse_frame}. \rev{The framework integrates two successive components, co-designed for computation and communication optimization.}
\begin{itemize}
\item To overcome the challenges of \textit{restricted onboard computation and satellite-GS connection in \S~\ref{subsec:limit_onboard_comp}}, \name deploys a regular LVLM at GS and a compact LVLM onboard, while proposing a progressive confidence network for efficient task allocation (\S~\ref{subsec:confidence_net}).
\item To overcome the challenges of \textit{high satellite data redundancy in \S~\ref{subsec:rs_redundancy}}, \name designs a text-image attention module to identify regions of interest, coupled with adaptive multi-scale processing to minimize transmission overhead (\S~\ref{subsec:multi_scale_preprocess}).
\end{itemize}

\subsection{Progressive Confidence Network}
\label{subsec:confidence_net}
In this section, we design a confidence network that leverages the progressively generated features of the onboard LVLM to determine whether a sample requires ground assistance.

\subsubsection{Problem Formulation}
To simplify the problem, we consider a single-satellite, single-GS scenario, which can be readily extended to multiple satellites and GS.
A compact LVLM, ${\bf{W}}^{\mathrm{s}}$, is deployed onboard, while a larger, more robust model, ${\bf{W}}^{\mathrm{g}}$, operates at GS for ground-based assistance ($|{\bf{W}}^{\mathrm{g}}|>|{\bf{W}}^{\mathrm{s}}|$).
Let ${\mathcal{D}} = \left\{ {{{\bf{x}}_{k}}, {\bf{T}}_k,{{\bf{y}}_{k}^{*}} } \right\}_{k = 1}^{{{\left| {{\mathcal D}} \right|}}}$ denote the set of input samples ${{\bf{x}}_{k}}$, text prompts ${\bf{T}}_k$, and their theoretical ground truths ${{\bf{y}}_{k}^{*}}$.
Let $\hat{\bf{y}}^{\mathrm{s}}_{k} = \phi({{\bf{x}}_{k}};{\bf{W}}^{\mathrm{s}})$ and $\hat{\bf{y}}^{\mathrm{g}}_{k} = \phi({{\bf{x}}_{k}};{\bf{W}}^{\mathrm{g}})$ denote the outputs of satellite- and GS-based LVLMs given input ${{\bf{x}}_{k}}$, where $\phi(\cdot; {\bf{W}})$ maps the input-output relationship of model ${\bf{W}}$.
Here, we omit the text prompt ${\bf{T}}_k$, as it remains unchanged between the satellite and GS for a given task.

The quality of output $\hat{\bf{y}}_{k}$ can be quantified by its similarity to the ground truth ${{\bf{y}}_{k}^{*}}$, denoted as $\text{Simi}(\hat{\bf{y}}_{k},{{\bf{y}}_{k}^{*}})$.
The metric $\text{Simi}(\cdot, \cdot)$ varies based on the evaluation criteria for each task (e.g., semantic similarity in question answering or equality in classification).
Clearly, the superior resources of GS support a larger 
model ${\bf{W}}^{\mathrm{g}}$, delivering better performance than the compact onboard model ${\bf{W}}^{\mathrm{s}}$, i.e., $\mathbb{E}_{({{\bf{x}}_{k}}, {\bf{T}}_k,{{\bf{y}}_{k}^{*}})\sim \mathcal{D}} \text{Simi}(\hat{\bf{y}}^{\mathrm{g}}_{k},{{\bf{y}}_{k}^{*}}) > \mathbb{E}_{({{\bf{x}}_{k}}, {\bf{T}}_k,{{\bf{y}}_{k}^{*}})\sim \mathcal{D}} \text{Simi}(\hat{\bf{y}}^{\mathrm{s}}_{k},{{\bf{y}}_{k}^{*}})$.
However, inference with ${\bf{W}}^{\mathrm{g}}$ introduces significant satellite-GS transmission latency.
To balance system latency and performance, we design an onboard confidence network to allocate tasks, determining whether each sample \( {{\bf{x}}_{k}} \) should be processed onboard to minimize latency or offloaded to GS for optimal performance.

\subsubsection{Network Design.}
Before onboard \( {\bf{W}}^{\mathrm{s}} \) inference, data ${{\bf{x}}_{k}} $ is encoded into latent features $V({{\bf{x}}_{k}})$ using a pretrained visual encoder \( V \).
We first build a simple neural network \( g \) that takes the image features $V({{\bf{x}}_{k}})$ as input to estimate the similarity between the satellite and ground LVLM outputs, $\text{Simi}(\hat{\bf{y}}^{\mathrm{s}}_{k},\hat{\bf{y}}^{\mathrm{g}}_{k})$.
A low similarity score indicates that the onboard output \( \hat{\bf{y}}^{\mathrm{s}}_{k} \) is likely unreliable, necessitating ground-based processing, whereas a high similarity suggests sufficient \textit{confidence} in the onboard inference.
Thus, optimizing the performance of \( g \) involves minimizing the discrepancy \( \mathbb{E} [ |g(V({{\bf{X}}_{k}})) - \text{Simi}(\hat{\bf{y}}^{\mathrm{s}}_{k},\hat{\bf{y}}^{\mathrm{g}}_{k}) | ] \) between the network's prediction and the true similarity.

\begin{figure}[t]
\centering
\includegraphics[width=.98\columnwidth]{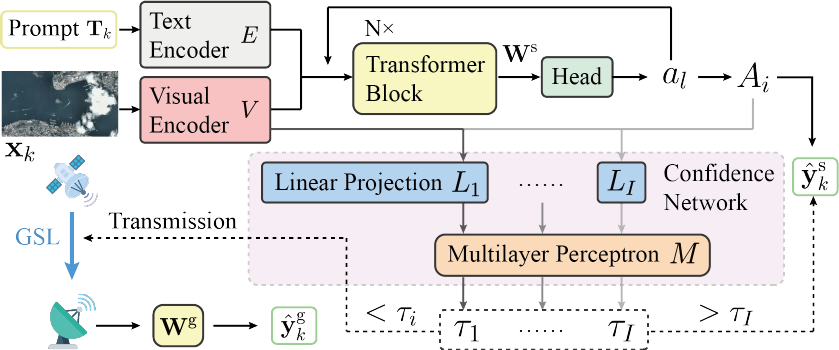}
\caption{The operational mechanism of the onboard progressive confidence network.
}
\label{fig:confidence_net}
\vspace{-1em}
\end{figure}

To further improve the fitting performance of \( g \), we incorporate \( \hat{\bf{y}}^{\mathrm{s}}_{k} \) as an additional input, redefining it as \( g^{\prime} \).
Compared to \( g(V({{\bf{X}}_{k}})) \), \( g^{\prime}(V({{\bf{X}}_{k}}),\hat{\bf{y}}^{\mathrm{s}}_{k}) \) incorporates \(\hat{\bf{y}}^{\mathrm{s}}_{k}\) as additional information, enabling more robust confidence estimation. However, this comes at the cost of increased latency, as transmission must await ${\bf{W}}^{\mathrm{s}}$ inference completion.
To harness the low-latency benefits of \( g \) while maintaining the robust performance of \( g^{\prime} \), we propose a progressive network architecture, \( \tilde{g} \), as shown in Figure~\ref{fig:confidence_net}.
Specifically, \( \tilde{g} \) consists of a multilayer perceptron $M$ equipped with \( I \) linear projections \( \{L_i\}_{i=1}^I \) in its initial layers, processing progressively received inputs with varying dimensions.

\subsubsection{Operational Workflow.}

For \( {{\bf{x}}_{k}} \), \( \tilde{g} \) first generates an initial confidence estimate (1-st iteration) sorely based on image features \( V({{\bf{x}}_{k}}) \), producing a similarity score \( \tilde{g}_1 (V({{\bf{x}}_{k}})) \).
Here, $\tilde{g}_1 = [L_1;M]$ consists of  \( M \) and projection \( L_1 \), with input dimensions aligned to $V({{\bf{x}}_{k}})$.
A threshold \( \tau_1 \) is defined such that if \( \tilde{g}_1 (V({{\bf{x}}_{k}})) < \tau_1 \), it indicates that \( V({{\bf{x}}_{k}}) \) alone provides sufficient confidence that ${\bf{W}}^{\mathrm{s}}$ is incapable of completing the task.
In this case, onboard inference is skipped, and the sample is directly transmitted to GS;
otherwise, ${\bf{W}}^{\mathrm{s}}$ inference continues to generate \( N_t \) additional tokens $a_i$, denoted as \( A_1 = \{a_l\}_{l=1}^{N_t} \), and perform a new confidence estimation \( \tilde{g}_2 (V({{\bf{x}}_{k}})),A_1) \).
Here, \( \tilde{g}_i = [L_i; M] \), where the input dimensions of \( L_i \) (for \( i > 1 \)) are aligned with the concatenated input \(\textrm{Concat}(V({{\bf{x}}_{k}}), A_{i-1})\).
Similarly, ${{\bf{x}}_{k}}$ for which \( \tilde{g}_i (V({{\bf{x}}_{k}}), A_{i-1}) \) falls below the threshold \( \tau_i \) are transmitted to the GS for further processing.
If not, ${\bf{W}}^{\mathrm{s}}$ generates the set of tokens \( A_{i} = \{a_l\}_{l=1}^{i \times N_t} \), followed by the next iteration $\tilde{g}_{i+1}$.


Until the final \( I \)-th iteration, once the complete output \( A_{I-1} = \hat{\bf{y}}^{\mathrm{s}}_{k} \) is obtained, \( \tilde{g} \) conducts a final confidence evaluation as \( \tilde{g}_I (V({{\bf{x}}_{k}}),A_{I-1}) \).
If \( \tilde{g}_I (V({{\bf{x}}_{k}}),A_{I-1}) < \tau_I \), \( {{\bf{x}}_{k}} \) is transmitted to GS for inference by ${\bf{W}}^{\mathrm{g}}$; otherwise, \( \hat{\bf{y}}^{\mathrm{s}}_{k} \) is output as the final result for \( {{\bf{x}}_{k}}  \).
Hereby, the progressive design of \( \tilde{g} \) integrate the advantages of both \( g \) and \( g^{\prime} \): i) for samples where early-stage inference is sufficiently confident that ${\bf{W}}^{\mathrm{g}}$ assistance is needed, \( \tilde{g} \) terminates ${\bf{W}}^{\mathrm{s}}$ early to conserve time and computational resources; ii) otherwise, \( \tilde{g}_i \) iterates, incorporating progressively generated tokens to enhance confidence estimation.

\subsubsection{Training Scheme.}

The parameter training of confidence network \( \tilde{g} \) follows a supervised learning approach.
The loss function for each sample ${{\bf{x}}_{k}} $ is defined as:
\begin{equation}
\mathcal{L}_k(\tilde{g}) = \sum_{i=1}^{I}\textrm{MSE} ( \tilde{g}_i (V({{\bf{x}}_{k}})),A_i) , \frac{ \hat{\bf{y}}^{\mathrm{s}}_{k} \cdot \hat{\bf{y}}^{\mathrm{g}}_{k}}{| \hat{\bf{y}}^{\mathrm{s}}_{k}||\hat{\bf{y}}^{\mathrm{g}}_{k}|} ),
\label{eq::loss_confidence}
\end{equation}
where $\textrm{MSE}$ denotes the mean square error.
$\mathcal{L}_k(\tilde{g})$ is designed to improve confidence assessment by employing a data-driven approach to update the parameters of \( \tilde{g} \), minimizing the discrepancy between the predicted and actual satellite-GS inference similarity at each iteration $\tilde{g}_i$.
Upon completing training, \( \tilde{g} \) aims to identify challenging samples for transmission to GS as early and accurately as possible. Its effectiveness will be thoroughly evaluated in the following experimental section.

\subsection{Attention-based Multi-scale Preprocessing}
\label{subsec:multi_scale_preprocess}

For \rev{satellite} image requiring satellite-GS transmission, we design an attention module to identify regions of interest, enabling adaptive preprocessing to reduce transmission overhead while maintaining inference accuracy.

\subsubsection{Problem Formulation}
\label{subsubsec::red_define}

We divide an input image \( \bf{x}_{k} \) into \( N^{\mathrm{r}}_{k} = \frac{H_k \times W_k}{H^{\mathrm{r}} \times W^{\mathrm{r}}} \) regions $\left\{ {{{\bf{x}}^r_{k}}} \right\}_{r = 1}^{{N^{\mathrm{r}}_{k} }}$, each with height \( H^{\mathrm{r}} \) and width \( W^{\mathrm{r}} \), where \( H_k \) and \( W_k \) denote the original size of $\bf{x}_{k}$.
Let \( {\tilde{\mathbf{x}}}_{k} = F(\mathbf{x}_{k})\) be processed by a redundancy filter \( F \), which preserves only ${\tilde{N}}^{\mathrm{r}}_{k}$ regions of interest while discarding the rest (${\tilde{N}}^{\mathrm{r}}_{k} \le {N}^{\mathrm{r}}_{k}$).
Redundancy filtering $F$ has a two-fold impact:
\begin{itemize}
  \item Removing certain regions may lead to the loss of global information, degrading inference performance, i.e., $\text{Simi}(\phi({{\bf{\tilde{x}}}}_{k};{\bf{W}}^{\mathrm{g}}),{{\bf{y}}_{k}^{*}}) < \text{Simi}(\phi({{\bf{x}}_{k}};{\bf{W}}^{\mathrm{g}}),{{\bf{y}}_{k}^{*}})$.
  \item The satellite-GS data transmission volume is reduced by ${ |{{{\bf{x}}_{k,f}}}|-|{\tilde{\mathbf{x}}}_{k,f} |}$, thereby lowering latency.
\end{itemize}
Therefore, the objective of \( F \) is to accurately evaluate the impact of each region on inference performance, aiming to minimize latency while preserving accuracy.

\subsubsection{Text-Image Attention.}

After processing by the onboard visual encoder \( V \), each region \( \left\{ {{{\bf{x}}^r_{k}}} \right\}_{r = 1}^{{N^{\mathrm{r}}_{k} }} \) is transformed into its corresponding image features $\left\{ V({{{\bf{x}}^r_{k}}} ) \right\}_{r = 1}^{{N^{\mathrm{r}}_{k}}}$.
Besides, the text prompt ${{T}_{k}}$ associated with ${{\bf{x}}_{k}}$ is processed by the text encoder \( E \) of ${\bf{W}}^{\mathrm{s}}$, which produces text features \( E(T_k) \).
At this stage, \( V({{{\bf{x}}^r_{k}}} ) \in \mathbb{R}^{N_V \times D_{{{\bf{W}}^{\mathrm{s}}}}} \) and \( E(T_k)  \in \mathbb{R}^{N_E \times D_{{{\bf{W}}^{\mathrm{s}}}}} \) reside in the same feature space within ${\bf{W}}^{\mathrm{s}}$, differing in token count ($N_V$ and $N_E$), while each token maintains an identical dimensionality (e.g., 1536 in Qwen2-VL-2B).
Accordingly, we compute the text-image attention~\cite{zhong2024urbancross} value for region ${{{\bf{x}}^r_{k}}}$:
\begin{equation}
K({{{\bf{x}}^r_{k}}}) =  \sum_{i=1}^{N_V} \sum_{j=1}^{N_E} \frac{V_i({{{\bf{x}}^r_{k}}}) \cdot E_j(T_k)}{||V_i({{{\bf{x}}^r_{k}}}) || || E_j(T_k)|| },
\label{eq::t_i_attention}
\end{equation}
with a higher value indicating greater relevance of the region to the given text prompt \( T_k \).
In the next section, we apply targeted preprocessing according to the attention scores \( K({{{\bf{x}}^r_{k}}}) \) to adaptively compress irrelevant regions, thus minimizing transmission overhead.

\RestyleAlgo{ruled}
\LinesNumbered
\begin{algorithm}[b]
\caption{The \name Worflow.}\label{alg:skyverse}
\setstretch{0.8}
\footnotesize
\SetKwInOut{Input}{Require}
\SetKwFunction{Fns}{Satellite-GS Task Allocation}
\SetKwFunction{Fg}{RS Redundancy Reduction}
\SetKwProg{Fn}{}{:}{}
\SetKwInOut{Output}{Output}

\Input{Visual encoder $V$, text encoder $E$, satellite LVLM ${\bf{W}}^{\mathrm{s}}$, GS LVLM ${\bf{W}}^{\mathrm{g}}$, confidence network $\tilde{g}$, and text-image attention thresholds $\alpha$ and $\beta$.}
\KwData{The set of input samples ${\mathcal{D}} = \left\{ {{{\bf{x}}_{k}}=\left\{ {{{\bf{x}}^r_{k}}} \right\}_{r = 1}^{{N^{\mathrm{r}}_{k} }}, {\bf{T}}_k} \right\}_{k = 1}^{{{\left| {{\mathcal D}} \right|}}}$.}
\Output{Inference results $\left\{\hat{\bf{y}}_{k}\right\}_{k = 1}^{{{\left| {{\mathcal D}} \right|}}}$.}

\For{each sample ${{\bf{x}}_{k}},{\bf{T}}_k$ in ${\mathcal{D}}$}{
\Fn{\Fns}{
    \For{each confidence estimate $i = 1,2,\ldots,I$}{
    \If{$ \tilde{g}_i (V({{\bf{x}}_{k}}),A_{i-1}) < \tau_i$}{
      ${\tilde{\mathbf{x}}}_{k} \leftarrow$ \textsf{RS Redundancy Reduction}$({{\bf{x}}_{k}},{\bf{T}}_k)$\;
      $\hat{\bf{y}}_{k} \leftarrow \phi({\tilde{\mathbf{x}}}_{k};{\bf{W}}^{\mathrm{g}})$ \quad\quad\quad\quad// \textit{GS Inference}\;
      }
    }
    $\hat{\bf{y}}_{k} \leftarrow \phi({{\mathbf{x}}}_{k};{\bf{W}}^{\mathrm{s}})$ \quad\quad\quad\quad// \textit{Satellite Inference}\;
}
}
\Fn{\Fg{${{\bf{x}}_{k}},{\bf{T}}_k$}}{
    \For{each RS region $r = 1,2,\ldots,N^{\mathrm{r}}_{k}$}{
    $K({{{\bf{x}}^r_{k}}}) \leftarrow  \sum\limits_{i=1}^{N_V} \sum\limits_{j=1}^{N_E} \frac{V_i({{{\bf{x}}^r_{k}}}) \cdot E_j(T_k)}{||V_i({{{\bf{x}}^r_{k}}}) || || E_j(T_k)|| }$ \quad// \textit{Attention Score}\;
    \If{$ K({{{\bf{x}}^r_{k}}}) < \alpha$}{
      ${\tilde{\bf{x}}^r_{k}} \leftarrow 0$ \quad\quad\quad\quad// \textit{Discarded Region}\;
      }
    \ElseIf{$\alpha \leq K({{{\bf{x}}^r_{k}}}) < \beta$}{
      ${\tilde{\bf{x}}^r_{k}} \leftarrow D({{{\bf{x}}^r_{k}}}, \frac{\beta - \alpha}{K({{{\bf{x}}^r_{k}}}) - \alpha})$ \quad\quad// \textit{Downsampled Region}\;
      }
    \Else{${\tilde{\bf{x}}^r_{k}} \leftarrow {{\bf{x}}^r_{k}}$ \quad\quad\quad\quad// \textit{Preserved Region}\;}
    }
    \KwRet $\left\{ {{\tilde{\bf{x}}^r_{k}}} \right\}_{r = 1}^{{N^{\mathrm{r}}_{k} }}$\;
}
\end{algorithm}

\begin{figure*}[t]
\centering
\includegraphics[width=.9\linewidth]{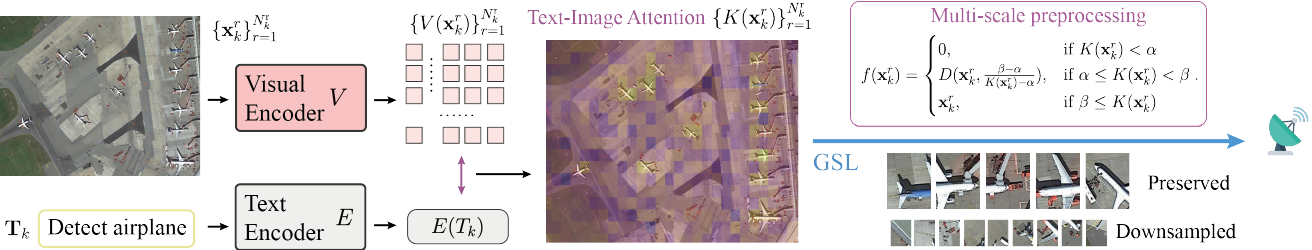}
\vspace{-.5em}
\caption{ 
\name adaptively compresses RS data by leveraging the attention between task prompt and different regions, minimizing satellite-GS transmission overhead while preserving inference performance.
}
\label{fig:image_text_attain}
\vspace{-1em}
\end{figure*}

\subsubsection{Multi-scale Preprocessing.}

Based on the relevance between image regions and the task prompt, we apply three processing distinct strategies: preserving the original resolution, downsampling, or discarding, as shown in Figure~\ref{fig:image_text_attain}.
Specifically, we design a multi-scale preprocessing function \( f \) as the redundancy filter \( F \) defined in Section~\ref{subsubsec::red_define}:
\begin{equation}
f({{{\bf{x}}^r_{k}}})=
\begin{cases} 
0, & \text{if } K({{{\bf{x}}^r_{k}}}) < \alpha \\
D({{{\bf{x}}^r_{k}}}, \frac{\beta - \alpha}{K({{{\bf{x}}^r_{k}}}) - \alpha}), & \text{if } \alpha \leq K({{{\bf{x}}^r_{k}}}) < \beta\\
{{{\bf{x}}^r_{k}}}, & \text{if } \beta \leq K({{{\bf{x}}^r_{k}}})
\end{cases}
.
\label{eq::f}
\end{equation}
Here, $D(x,c)$ denotes the downsampling operation applied to the original image region 
$x$ based on the scaling factor $c$, and \( \alpha \) and \( \beta \) serve as threshold values for text-image attention scores.
Regions with \( K({{{\bf{x}}^r_{k}}}) < \alpha \) are deemed largely irrelevant to the current task and have minimal impact on inference performance. To minimize communication overhead, \name discards these regions instead of transmitting them to GS.
When \( K({{{\bf{x}}^r_{k}}})  > \beta \), it indicates that the region \( {{{\bf{x}}^r_{k}}} \) is important for the current inference task. In this case, \name retains its original resolution to ensure sufficient information is transmitted.
Moreover, for regions with attention values between \(\alpha\) and \(\beta\), \( f \) applies downsampling to strike a balance between information preservation and transmission cost.
The scaling factor of downsampling is set to $\frac{\beta - \alpha}{K({{{\bf{x}}^r_{k}}}) - \alpha}$ and decreases as the region's importance increases. Once the importance surpasses \( \beta \), the scaling factor reaches 1, meaning \name retains this region at its original resolution without compression.

In summary, for Earth observation tasks, \name first utilizes an onboard confidence network to progressively determine whether a sample requires ground processing. Before initiating satellite-GS transmission, multi-scale preprocessing is applied to reduce satellite data redundancy. The complete pseudocode of the \name workflow is presented in Algorithm~\ref{alg:skyverse}.

\section{Evaluation}
\label{sec:eval}


\subsection{Implementation}
\label{subsec:implement}

\subsubsection{Prototypes}
\begin{figure}[b]
\vspace{-1em}
\centering
\includegraphics[width=.925 \columnwidth]{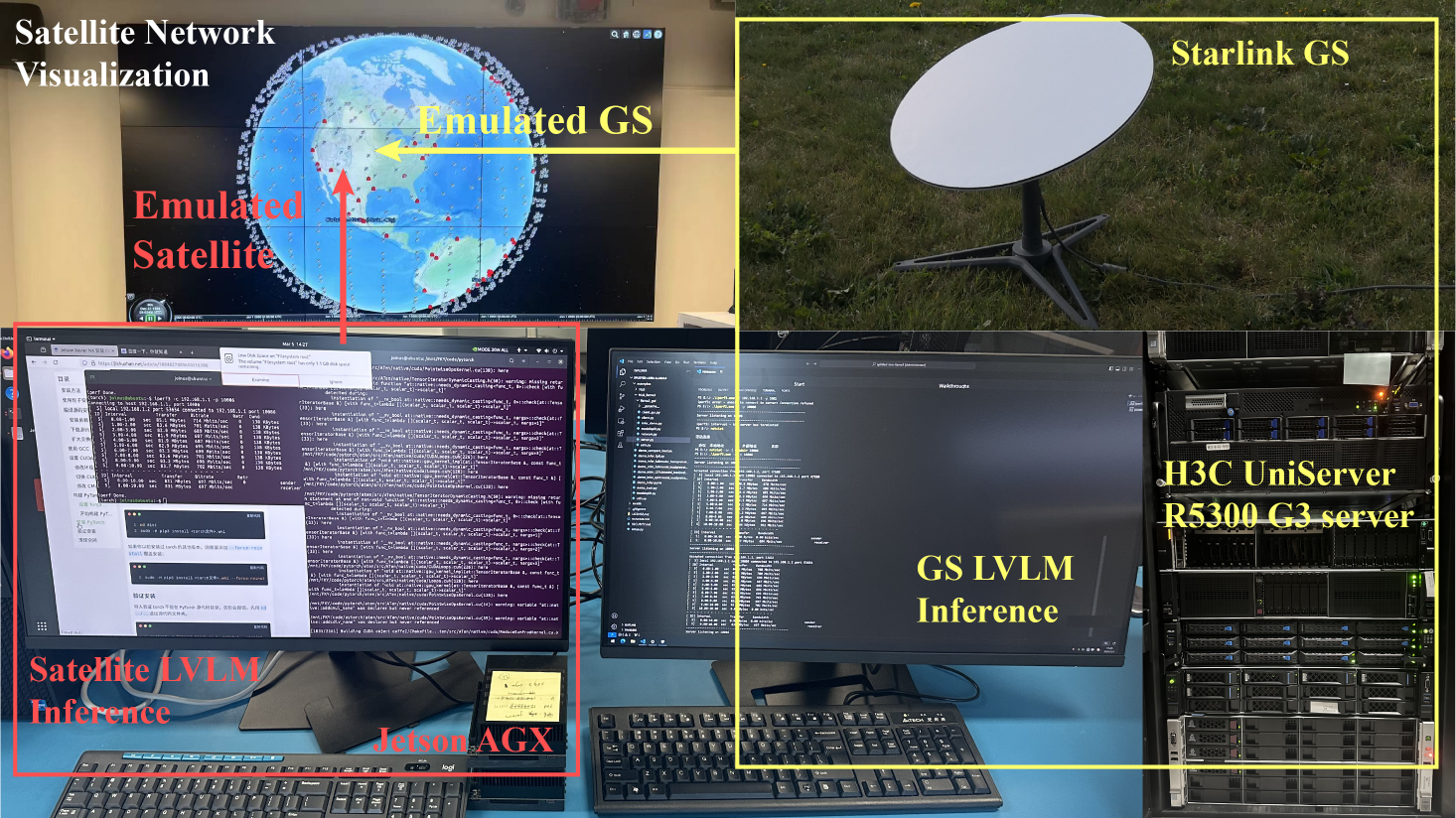}
\vspace{-1em}
\caption{ 
\name prototype and testbed.
}
\label{fig:implementation}
\end{figure}

We develop the \name prototype for satellite-GS communication and LVLM inference, as depicted in Figure~\ref{fig:implementation}.
The GS LVLM is deployed on an H3C UniServer R5300 G3 server, featuring eight NVIDIA GeForce RTX 3090 GPUs, dual Intel Xeon Silver 4210R processors (10 cores, 2.84 GHz each), and 8×32 GB DDR4 RAM.
The satellite LVLM is deployed on the Jetson AGX Xavier, featuring an 8-core ARM v8.2 64-bit CPU, 16 GB EMMC 5.1 storage.
The software stack includes Ubuntu 18.04.6 LTS, Python 3.7 and PyTorch 1.9.1.
The satellite-GS link is emulated and configured using \textit{Open vSwitch}~\cite{pfaff2015design}, with traffic conditions controlled by \textit{tc}~\cite{beshay2015fidelity} based on Starlink traces collected by our commercial Starlink GS.
\rev{Real-world constellations~\cite{starlink_gp} (e.g., two-line element) of Starlink and Planet are incorporated to enable real-time trajectory computation, establishing links between satellites and GS.} A 3D interface visualizes these connections and data exchanges, closely replicating real-world satellite communication dynamics.

\subsubsection{RS Tasks}

\begin{figure*}[t] 
    \setlength\abovecaptionskip{6pt}
     \setlength\subfigcapskip{0pt}
        \subfigure[{RSVQA.}]{	
        \centering
		\label{subfig:overall_rsvqa}
		\includegraphics[width=.28\linewidth]{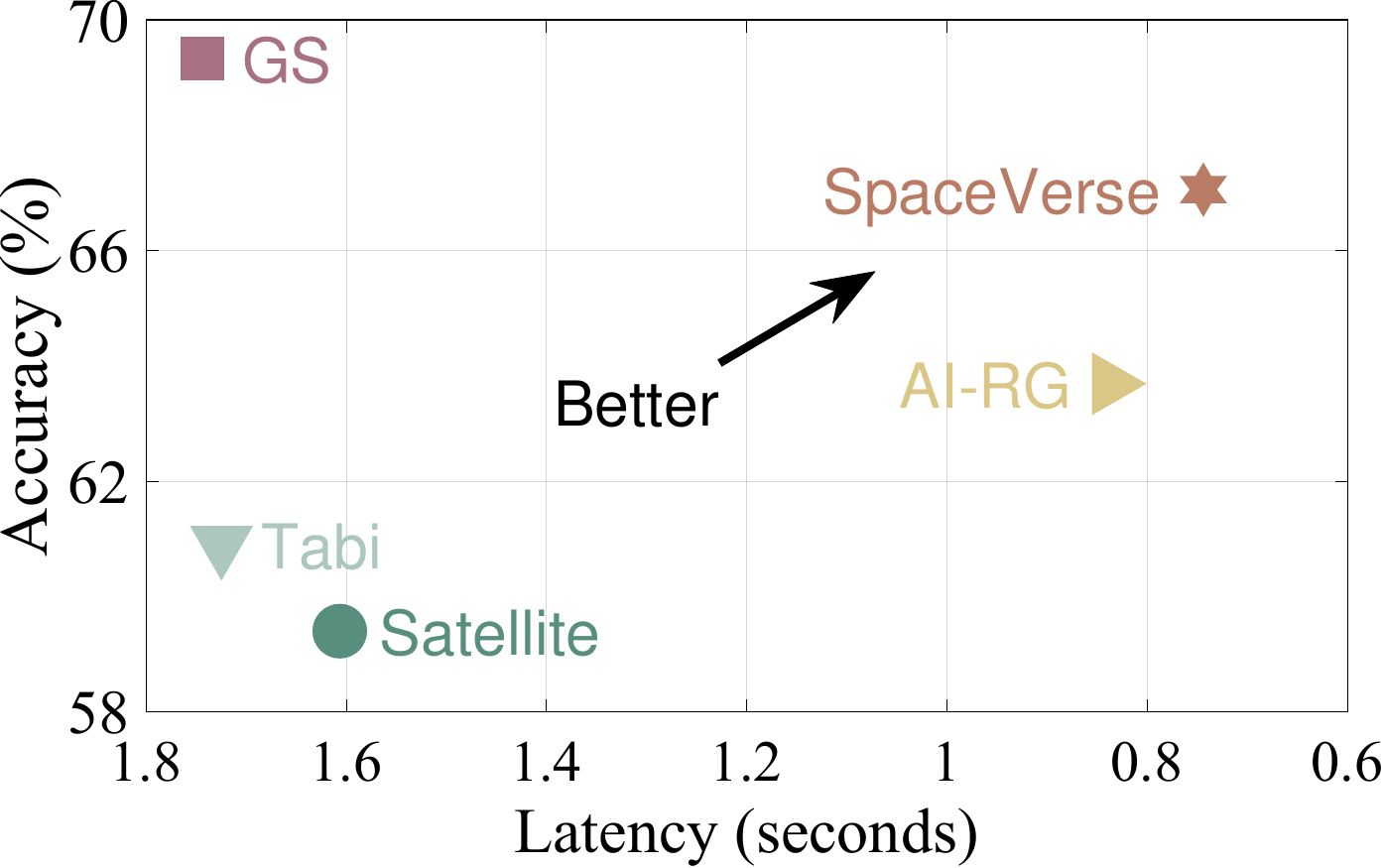}}
    \hspace{.5em}
	\subfigure[{RESISC.}]{
        \centering
		\label{subfig:overall_resisc}
		\includegraphics[width=.277\linewidth]{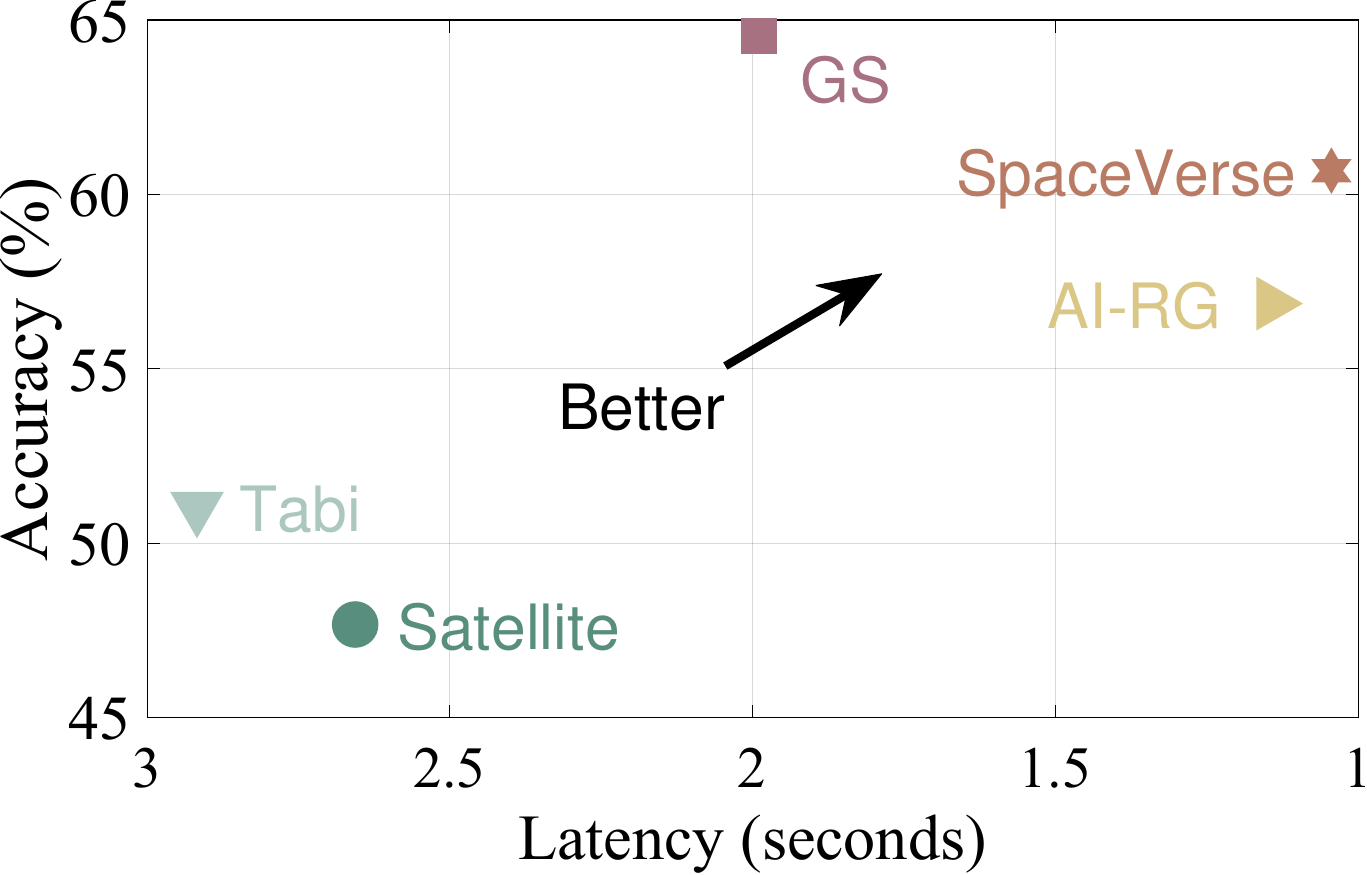}}
    \hspace{.5em}
        \subfigure[{DOTA.}]{
        \centering
		\label{subfig:overall_dota}
		\includegraphics[width=.29\linewidth]{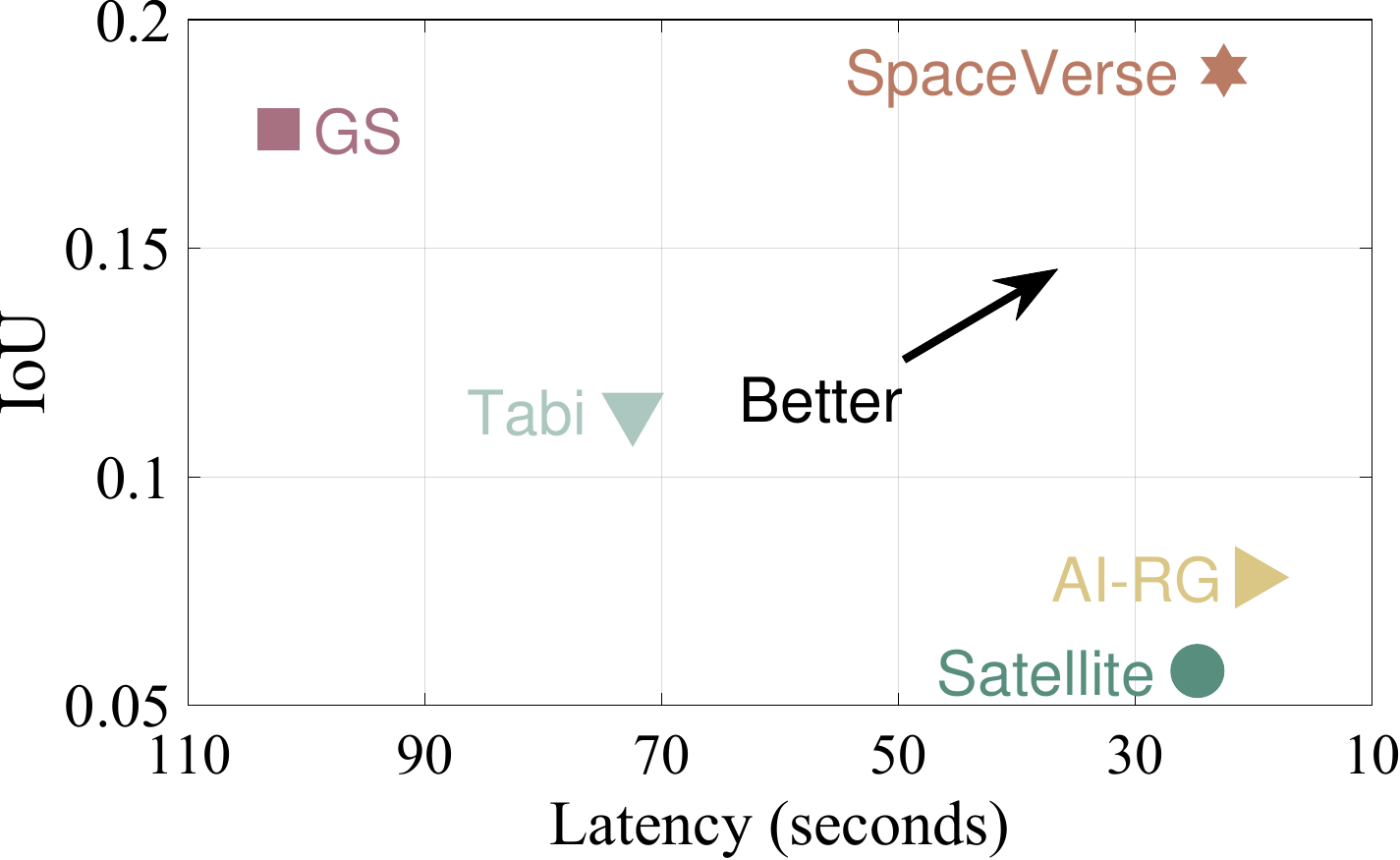}}
\vspace{-.5em}
\caption{
The inference latency and performance on RSVQA, RESISC, and DOTA datasets.
}
    \label{fig:overall_exp}
\vspace{-1.5em}
\end{figure*}

We use three representative RS tasks: visual question answering, image classification, and object detection, using the RSVQA LR~\cite{lobry2020rsvqa}, RESISC45~\cite{cheng2017remote}, and DOTA-v1.0~\cite{xia2018dota} datasets.
We evaluate the inference time and performance per sample in test sets, with 5\% train sets used to update the parameters of \( \tilde{g} \).
RSVQA LR, derived from Sentinel-2 imagery over Netherlands,  comprises 772 images (256×256 pixels, 10m resolution) with RGB bands, and we evaluate 3,500 questions across presence and comparison tasks.
RESISC45 comprises 31,500 images across 45 scene classes, including airplanes, bridges, and islands. Each image is 256×256 pixels, with spatial resolutions ranging from 30m to 0.2m per pixel.
DOTA-v1.0 contains images from diverse sensors and platforms, with sizes ranging from 800×800 to 20,000×20,000 pixels. We evaluate six target categories: ground-track field, soccer field, swimming pool, baseball diamond, basketball court, and tennis court.

\subsubsection{Model}

We employ the well-known Qwen2-VL series~\cite{wang2024qwen2}, a family of LVLM designed for comprehensive image understanding, including image captioning, visual question answering, and grounded reasoning. We deploy Qwen2-VL 7B on GS, while the compact 2B model runs on the satellite.
We adopt the widely used CLIP-ViT-Patch/16 model~\cite{radford2021learning} for embedding text-image attention.

\subsubsection{Hyperparameters}

The granularity of multi-scale preprocessing $N^{\mathrm{r}}_{k}$ is set to 100.
The satellite-to-GS bandwidth is set to 110.67 Mbps, based on measurements from our Starlink GS.
We select 10 satellites from the 570km Starlink constellation and evenly distribute the test data across them.
$I$ is set to 2, with $\tau_1=0.5$ and $\tau_2=0.4$, and the values of $\alpha$ and $\beta$ are set to 0.35 and 0.55, respectively.

\subsubsection{Benchmarks}

\begin{figure}[t] 
    \setlength\abovecaptionskip{6pt}
     \setlength\subfigcapskip{0pt}
        \subfigure[{RSVQA.}]{	
        \centering
		\label{subfig:varyofffload_rsvqa}
		\includegraphics[width=.475\columnwidth]{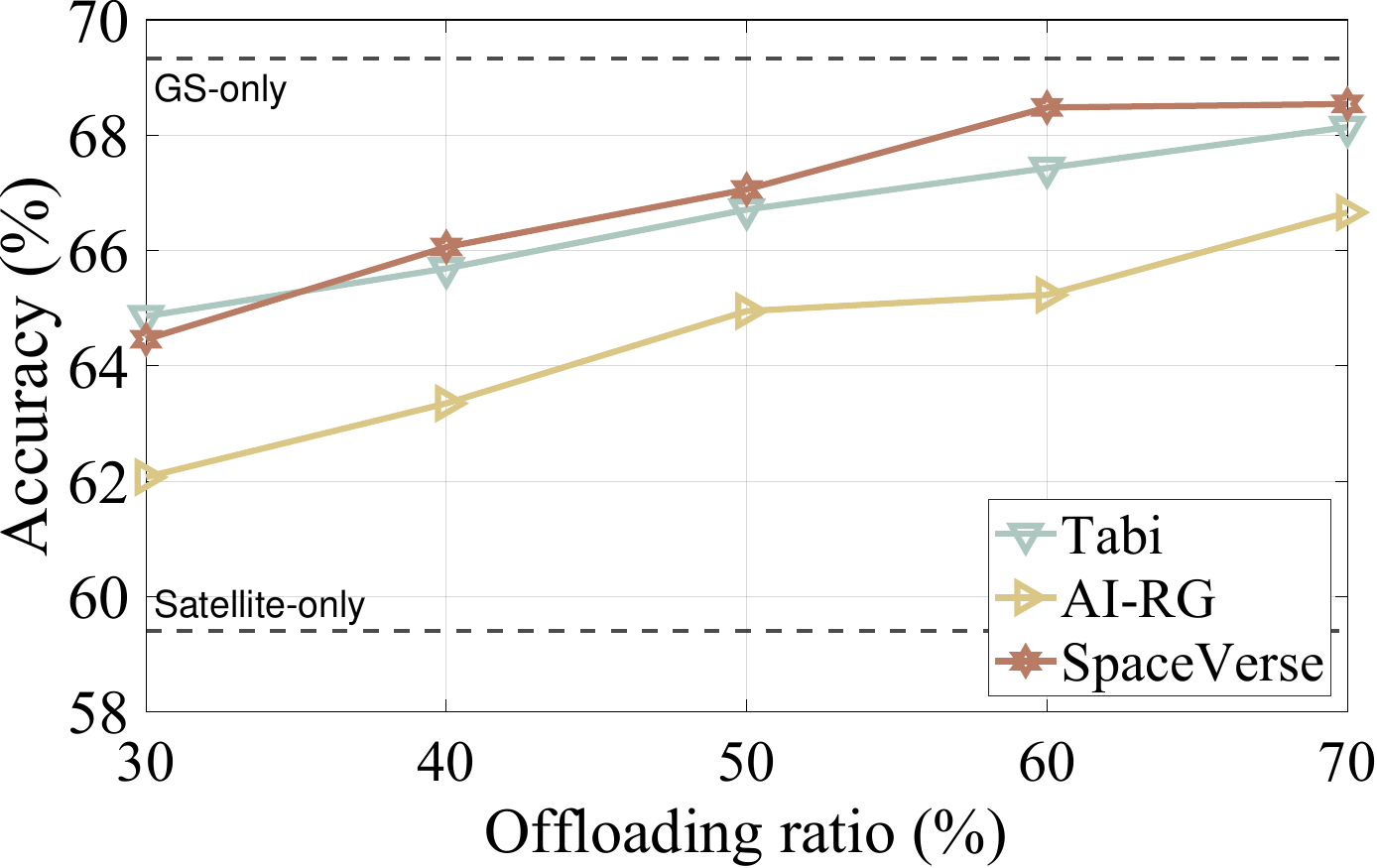}}
	\subfigure[{RESISC.}]{
        \centering
		\label{subfig:varyofffload_resisc}
		\includegraphics[width=.475\columnwidth]{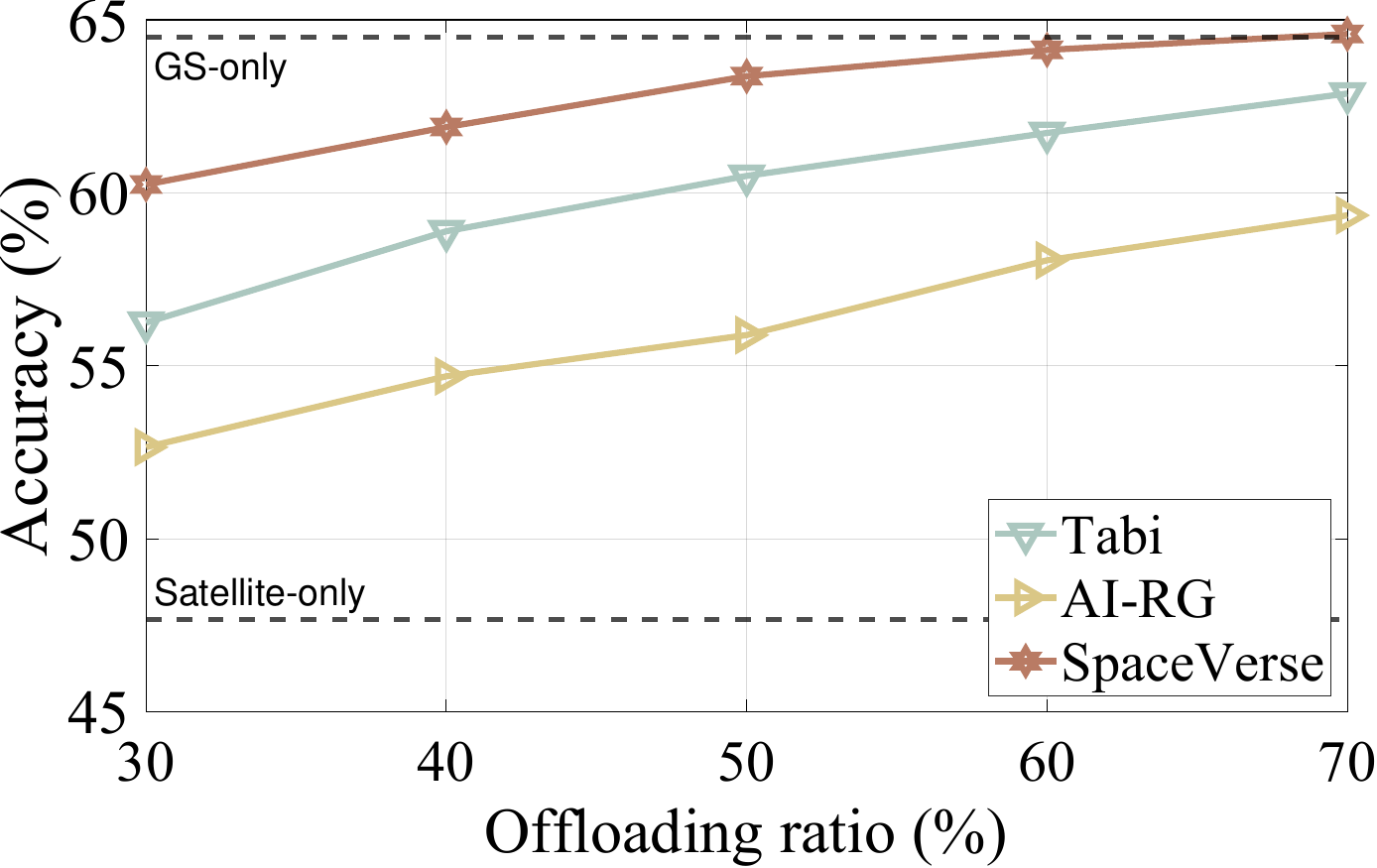}}
\vspace{-.5em}
\caption{
The impact of offloading volume on system performance on RSVQA and RESISC datasets.
}
    \label{fig:impact_offload}
\vspace{-1.5em}
\end{figure}

We compare \name with:
1) the status-quo satellite- and GS-only baselines and 2) the two state-of-the-art edge-cloud synergistic model inference systems Tabi{\cite{wang2023tabi}} and AI-RG (active inference with rewardless guidance offloading strategy){\cite{he2024large}}.
Tabi employs a single confidence score for task allocation, using word pruning and weighted ensemble techniques to mitigate system overhead and accuracy loss.
AI-RG leverages active inference to optimize and balance edge-cloud task offloading and resource allocation in synergistic inference.

\subsection{Overall Performance}
\label{subsec:over_all}

Figure~\ref{fig:overall_exp} presents a comprehensive comparison of \name with benchmarks in terms of average per-sample inference latency and performance.
The satellite-only approach exhibits the lowest performance among all methods, as its limited resources can only support a constrained 2B model.
In contrast, the GS-only approach achieves strong performance through 7B model, but suffers from severely increased latency (up to 4.14$\times$ on the DOTA dataset) due to the transmission of raw RS data over constrained satellite-GS links.
Tabi offloads samples to GS based on onboard confidence, achieving an average performance gain of {36.2\%} compated to satellite LVLM. However, each offloaded sample requires complete onboard inference beforehand, resulting in substantial time overhead by {69.9\%}.
In comparison, AI-RG incurs only 58.7\% of the onboard latency overhead, due to its activate inference strategy that jointly optimizes computation and communication. However, without a performance-aware offloading policy, its performance reaches only 74.8\% of that achieved by GS model.
Thanks to accurate task allocation and effective redundancy reduction, \name outperforms other benchmarks by an average of {31.2\%} in performance and reduces latency by {51.2\%}, achieving overall optimal results.

\begin{figure}[t] 
    \setlength\abovecaptionskip{6pt}
     \setlength\subfigcapskip{0pt}
        \subfigure[{RSVQA.}]{	
        \centering
		\label{subfig:abla_conf_rsvqa}
		\includegraphics[width=.475\columnwidth]{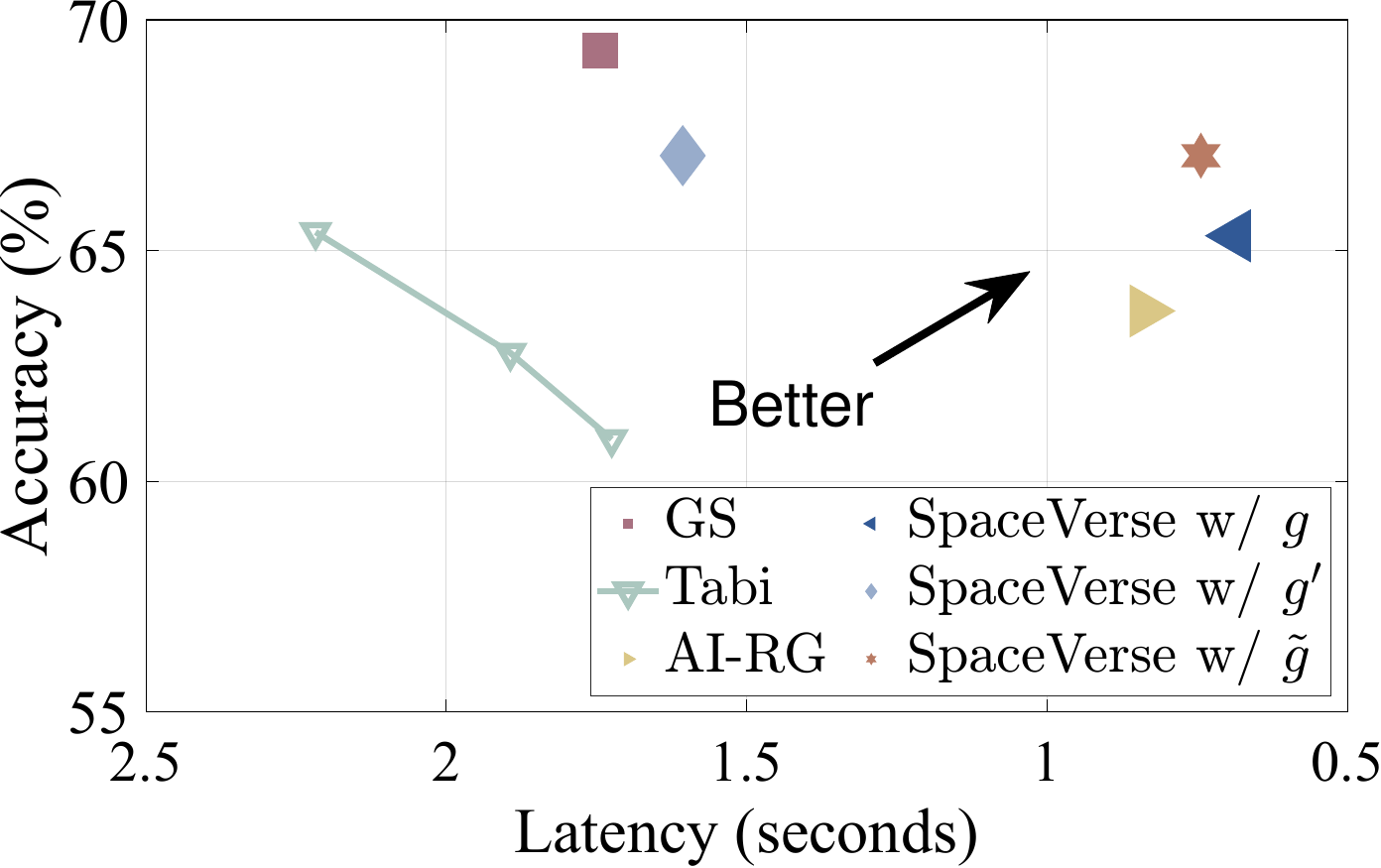}}
	\subfigure[{DOTA.}]{
        \centering
		\label{subfig:abla_conf_dota}
		\includegraphics[width=.475\columnwidth]{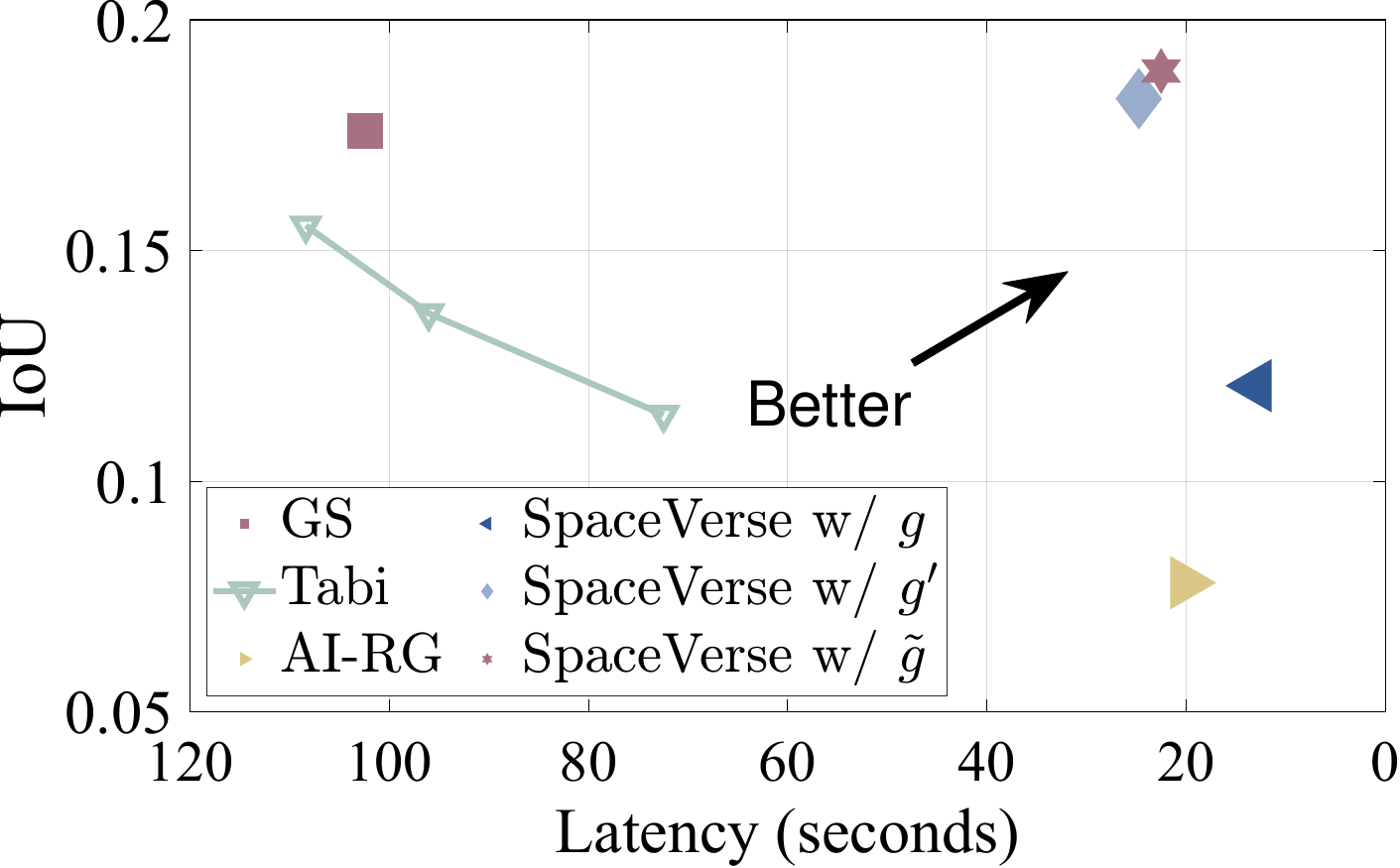}}
\vspace{-.5em}
\caption{
Ablation experiments for progressive confidence network on RSVQA and DOTA datasets.
}
    \label{fig:abla_conf}
\vspace{-1.5em}
\end{figure}

\subsection{The Impact of Offloading Volume}
\label{subsec:bandwidth_impact}

Figure~\ref{fig:impact_offload} shows the impact of the number of samples offloaded to GS on inference performance.
As the number of offloaded samples increases, all methods show improved performance, driven by the use of more powerful LVLM enabled by high-performance ground computing.
Under identical offloading conditions, AI-RG performs the worst, as its selection ignores sample difficulty, resulting in unnecessary transmission of easy samples.
Tabi achieves notably improved task allocation performance by leveraging confidence scores derived from output token probabilities.
In comparison, by incorporating both image features and onboard outputs into a neural network-based decision module, \name enables more effective task allocation, yielding an average performance gain of {6.2\%} over other benchmarks.


\subsection{Ablation Study}
\label{subsec:ablation}

\subsubsection{Progressive Confidence Network}

Figure~\ref{fig:abla_conf} shows the effectiveness
of progressive confidence network.
Tabi adjusts its confidence threshold to enable a tunable trade-off between performance and latency.
Evidently, \rev{the carefully designed \name achieves optimal accuracy–latency trade-offs through progressive task allocation strategy.}
Using only \( g \) yields the lowest latency, but its limited input scope compromises allocation quality, resulting in performance degradation.
In contrast, \( g^{\prime} \) achieves better performance, but the need for full onboard LVLM inference leads to a significant increase in latency.
\name with \( \tilde{g} \) leverages \( g \) to promptly offload difficult samples, reducing onboard computation, and use \( g^{\prime}  \) to ensure optimal allocation, thus achieving strong performance with minimal system latency.

\begin{figure}[t] 
    \setlength\abovecaptionskip{6pt}
     \setlength\subfigcapskip{0pt}
        \subfigure[RESISC]{	
        \centering
		\label{subfig:abla_redund_resisc}
		\includegraphics[width=.475\columnwidth]{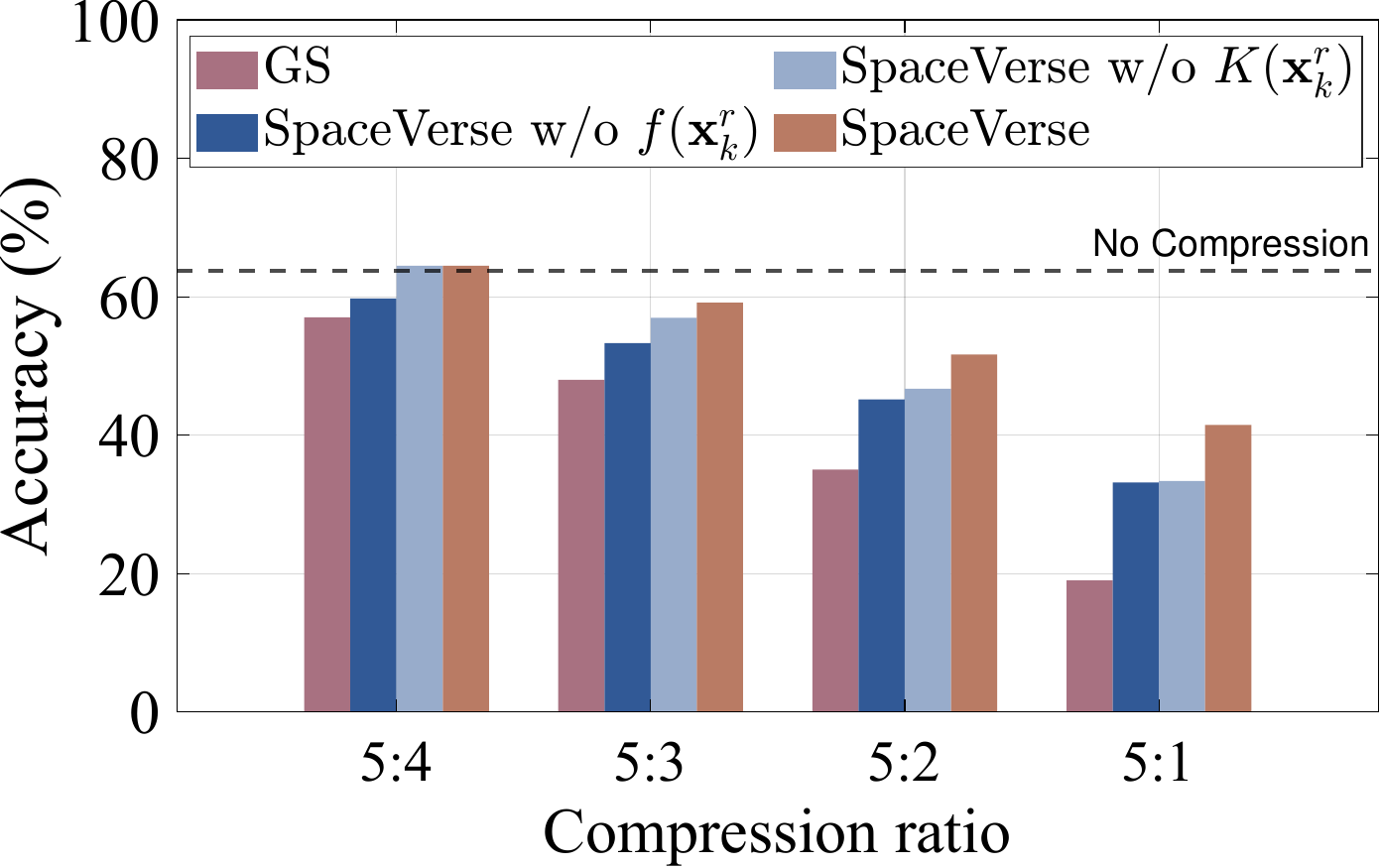}}
	\subfigure[DOTA]{
        \centering
		\label{subfig:abla_redund_dota}
		\includegraphics[width=.475\columnwidth]{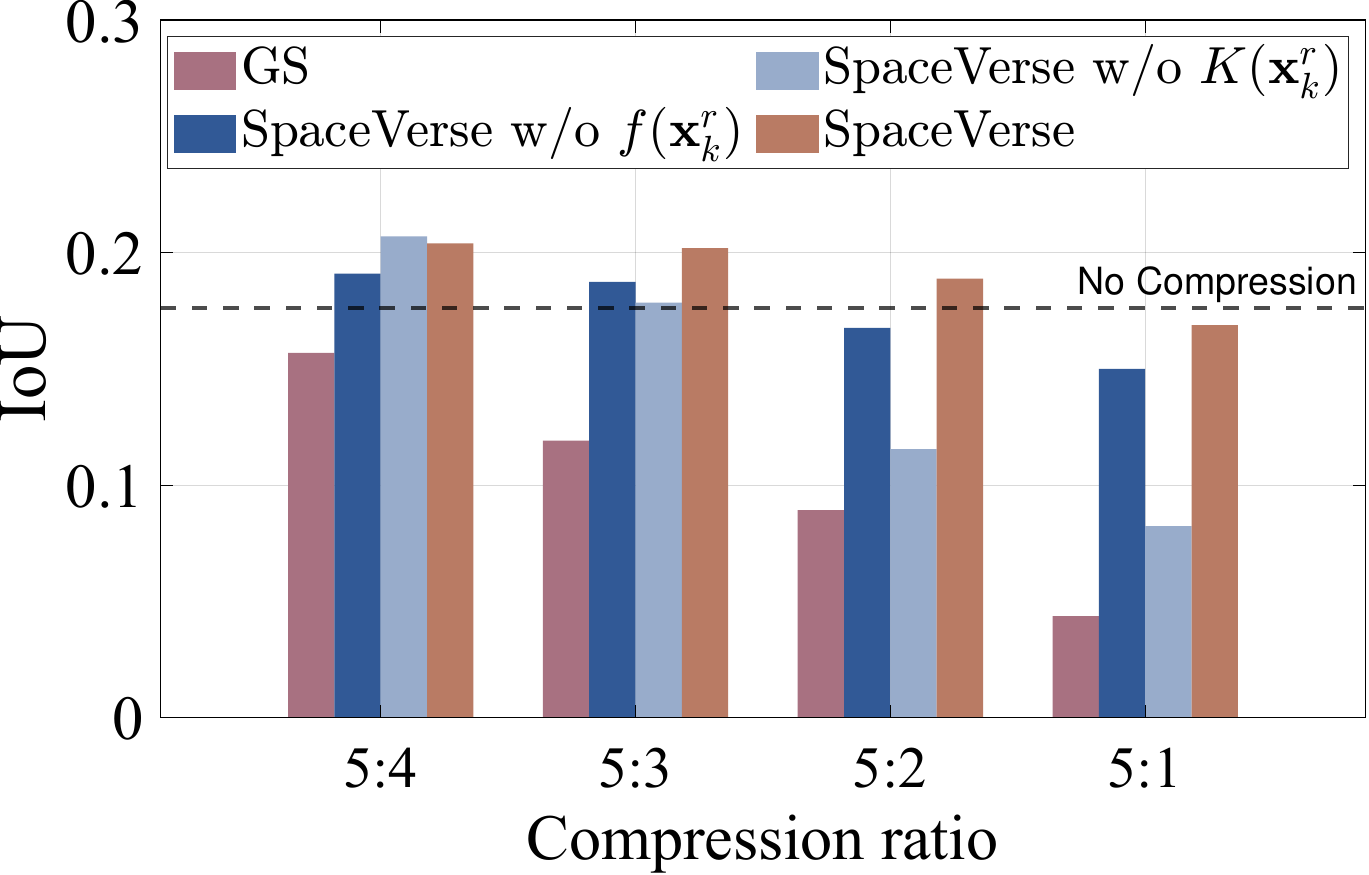}}
    \vspace{-.5em}
    \\
    \subfigure[Satellite-to-GS transmission visualization]{
        \centering
		\label{subfig:abla_redund_dota}
		\includegraphics[width=.99\columnwidth]{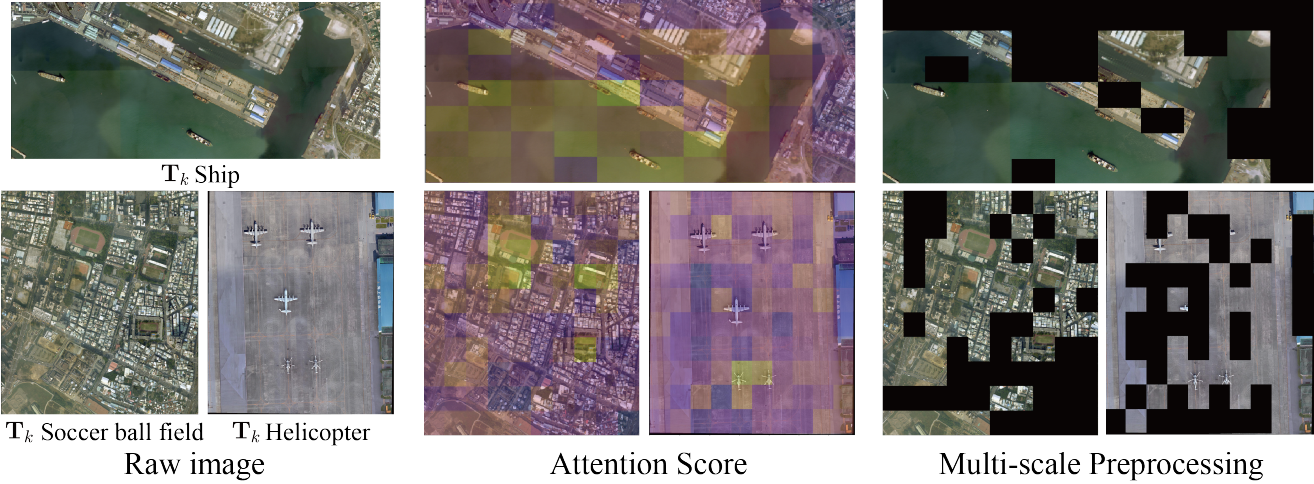}}
\vspace{-.5em}
\caption{
Ablation studies and visualization for multi-scale preprocessing on RESISC and DOTA datasets.
}
    \label{fig:abla_redund}
\vspace{-1.5em}
\end{figure}

\subsubsection{Attention-based Multi-scale
Preprocessing}
Figure~\ref{fig:abla_redund} shows the performance of ablated variants in different transmission compression ratios.
As the compression ratio increases, all methods experience varying degrees of performance degradation. GS-only, which employs a random masking strategy, suffers the most—exhibiting {72.7\%} performance drop at a 5:1 compression ratio.
By contrast, text-image attention $K({{{\bf{x}}^r_{k}}}) $ and multi-scale preprocessing $f({{{\bf{x}}^r_{k}}}) $ enhance robustness by adaptively identifying critical regions and flexibly preserving informative content, respectively.
\name achieves optimal performance under high compression ratios through the integration of both designs, with only {4.1\%} performance drop compared to uncompressed settings on DOTA dataset, demonstrating remarkable robustness.

\section{Related Work and Discussion}
\label{sec:related_work}

\paragraph{LVLMs for Earth observation}
\label{subsec:related_lvlm_for_rs}

Transformer-based LVLMs have demonstrated robust capabilities in comprehensive image understanding, driving research into their application in Earth observation.
Hu~\textit{et al.}{~\cite{hu2023rsgpt}} develop RSICap, a satellite image captioning dataset, and utilize it to train an LVLM specifically designed for Earth observation.
Kuckreja~\textit{et al.}{~\cite{kuckreja2024geochat}} develop an LVLM, enabling multitask dialogue with a focus on region-level reasoning and visual localization in high-resolution satellite imagery.
Zhang~\textit{et al.}{~\cite{zhang2024earthgpt}} propose MMRS-1M, a large-scale satellite expert knowledge dataset, and integrate a vision-enhanced perception mechanism into LVLMs, combining coarse-grained semantics with fine-grained details.
However, existing studies primarily emphasize model design and training, while overlooking challenges in real-world deployment. LVLMs demand computational resources far exceeding satellites' capabilities, while the high volume of high-resolution data can overwhelm the limited satellite-GS bandwidth, posing significant obstacles to near real-time deployment of LVLM in Earth observation.

\paragraph{Edge-Cloud Synergistic Inference}
\label{subsec:related_syn_infer}

Edge-cloud synergistic inference harnesses the cloud's robust computational power while capitalizing on the low-latency advantages of edge computing~\cite{yao2022edge,lin2024splitlora,zhang2024fedac}.
He~\textit{et al.}{~\cite{he2024large}} propose an active inference approach to optimize task offloading and resource allocation, overcoming limitations in data efficiency, latency sensitivity, and adaptability to task load variations in edge-cloud synergistic systems.
Borzunov~\textit{et al.}{~\cite{borzunov2022petals}} propose a split inference framework that partitions large AI models between edge and cloud, enabling synergistic inference while coordinating computational resources across multiple nodes.
Wang~\textit{et al.}{~\cite{wang2023tabi}} design a multi-tier cascading model service system that integrates early-exit mechanisms, attention-based word pruning, and weighted multi-level ensemble learning, reducing the latency and cost of edge-cloud synergistic inference.
Despite these advancements, deploying edge-cloud synergistic LVLM inference in LEO satellite networks remains challenging. Satellite-GS links suffer from lower bandwidth and stability compared to terrestrial networks, while satellite data exhibit high redundancy in LVLM inference, making current methods inefficient. Research on satellite-GS synergistic LVLM inference remains largely unexplored.

\section{Conclusion}
\label{sec:conclusion}

\rev{Taking a significant step toward advanced Earth observation, we propose \name, a satellite-ground synergistic LVLM inference system designed for LEO satellite networks.}
\name consists of two primary components: progressive confidence network and multi-scale satellite data preprocessing.
First, the progressive confidence network optimally allocates tasks between the onboard compact LVLM and its ground-based counterpart, balancing \rev{accuracy and latency} under satellite computation and communication constraints.
Second, \rev{a text-image attention-based multi-scale preprocessing mechanism reduces data redundancy during satellite-GS transmission, further improving system efficiency while preserving accuracy.}
Extensive experiments on real-world datasets and LEO satellite constellations show that \name outperforms state-of-the-art benchmarks.

\balance

\bibliographystyle{unsrt}
\bibliography{reference}

\begin{thebibliography}{10}

\bibitem{liu2024democratizing}
Lixin Liu, Yuanjie Li, Hewu Li, Jiabo Yang, Wei Liu, Jingyi Lan, Yufeng Wang, Jiarui Li, Jianping Wu, Qian Wu, et~al.
\newblock {Democratizing $\{$Direct-to-Cell$\}$ Low Earth Orbit Satellite Networks}.
\newblock In {\em Proc. of the 21st NSDI}, pages 791--808, 2024.

\bibitem{fang2024robust}
Hao Fang, Haoyuan Zhao, Jianxin Shi, Miao Zhang, Guanzhen Wu, Yi~Ching Chou, Feng Wang, and Jiangchuan Liu.
\newblock {Robust Live Streaming over LEO Satellite Constellations: Measurement, Analysis, and Handover-Aware Adaptation}.
\newblock In {\em Proc. of the 32nd ACM MM}, pages 5958--5966, 2024.

\bibitem{singh2024spectrumize}
Vaibhav Singh, Tusher Chakraborty, Suraj Jog, Om~Chabra, Deepak Vasisht, and Ranveer Chandra.
\newblock {Spectrumize: Spectrum-Efficient Satellite Networks for the Internet of Things}.
\newblock In {\em Proc. of the 21st NSDI}, pages 825--840, 2024.

\bibitem{zhang2024satfed}
Yuxin Zhang, Zheng Lin, Zhe Chen, Zihan Fang, Wenjun Zhu, Xianhao Chen, Jin Zhao, and Yue Gao.
\newblock {SatFed: A Resource-Efficient LEO Satellite-Assisted Heterogeneous Federated Learning Framework}.
\newblock {\em arXiv preprint arXiv:2409.13503}, 2024.

\bibitem{lin2025leo}
Zheng Lin, Yuxin Zhang, Zhe Chen, Zihan Fang, Cong Wu, Xianhao Chen, Yue Gao, and Jun Luo.
\newblock {LEO-Split: A Semi-Supervised Split Learning Framework over LEO Satellite Networks}.
\newblock {\em arXiv preprint arXiv:2501.01293}, 2025.

\bibitem{lin2025esl}
Zheng Lin, Yuxin Zhang, Zhe Chen, Zihan Fang, Yanni Yang, Guoming Zhang, Huan Yang, Cong Wu, Xianhao Chen, and Yue Gao.
\newblock {ESL-LEO: An Efficient Split Learning Framework over LEO Satellite Networks}.
\newblock In {\em Proc. Int. Conf. Wireless Artif. Intell. Comput. Syst. Appl.}, pages 344--357, 2025.

\bibitem{ahmmed2022digital}
Tuheen Ahmmed, Afsoon Alidadi, Zichao Zhang, Aizaz~U Chaudhry, and Halim Yanikomeroglu.
\newblock {The Digital Divide in Canada and the Role of LEO Satellites in Bridging the Gap}.
\newblock {\em {IEEE} Commun. Mag.}, 60(6):24--30, 2022.

\bibitem{planet_lab}
``{{Planet}}", 2024.
\newblock Available: https://www.planet.com/.

\bibitem{yuan2024satsense}
Haoxuan Yuan, Zhe Chen, Zheng Lin, Jinbo Peng, Zihan Fang, Yuhang Zhong, Zihang Song, and Yue Gao.
\newblock {SatSense: Multi-Satellite Collaborative Framework for Spectrum Sensing}.
\newblock {\em {IEEE} Trans. Cogn. Commun. Netw.}, 2025.

\bibitem{peng2025sigchord}
Jinbo Peng, Junwen Duan, Zheng Lin, Haoxuan Yuan, Yue Gao, and Zhe Chen.
\newblock {SigChord: Sniffing Wide Non-Sparse Multiband Signals for Terrestrial and Non-Terrestrial Wireless Networks}.
\newblock {\em arXiv preprint arXiv:2504.06587}, 2025.

\bibitem{lin2024fedsn}
Zheng Lin, Zhe Chen, Zihan Fang, Xianhao Chen, Xiong Wang, and Yue Gao.
\newblock {FedSN: A Federated Learning Framework over Heterogeneous LEO Satellite Networks}.
\newblock {\em {IEEE} Trans. Mobile Comput.}, 2024.

\bibitem{zhao2024leo}
Zhiyuan Zhao, Zhe Chen, Zheng Lin, Wenjun Zhu, Kun Qiu, Chaoqun You, and Yue Gao.
\newblock {LEO Satellite Networks Assisted Geo-Distributed Data Processing}.
\newblock {\em {IEEE} Wireless Commun. Lett.}, 2024.

\bibitem{li2023networking}
Yuanjie Li, Hewu Li, Wei Liu, Lixin Liu, Wei Zhao, Yimei Chen, Jianping Wu, Qian Wu, Jun Liu, Zeqi Lai, et~al.
\newblock {A Networking Perspective on Starlink's Self-Driving LEO Mega-Constellation}.
\newblock In {\em Proc. of the 29th ACM MobiCom}, pages 1--16, 2023.

\bibitem{shenoy2024s4}
Jayanth Shenoy, Xinjian~Davis Zhang, Shlok Mehrotra, Bill Tao, Rem Yang, Han Zhao, and Deepak Vasisht.
\newblock {S4: Self-Supervised Sensing Across the Spectrum}.
\newblock {\em arXiv preprint arXiv:2405.01656}, 2024.

\bibitem{hu2024utilizing}
Wenmiao Hu.
\newblock {Utilizing Very High-resolution Optical RGB Satellite Imagery in Geo-information Extraction for Fine-scale Map-making}.
\newblock In {\em Proc. of the 32nd ACM MM}, pages 11127--11131, 2024.

\bibitem{dewitte2021artificial}
Steven Dewitte, Jan~P Cornelis, Richard M{\"u}ller, and Adrian Munteanu.
\newblock {Artificial Intelligence Revolutionises Weather Forecast, Climate Monitoring and Decadal Prediction}.
\newblock {\em Remote Sensing}, 13(16):3209, 2021.

\bibitem{bhaga2020impacts}
Trisha~Deevia Bhaga, Timothy Dube, Munyaradzi~Davis Shekede, and Cletah Shoko.
\newblock {Impacts of Climate Variability and Drought on Surface Water Resources in Sub-Saharan Africa Using Remote Sensing: A Review}.
\newblock {\em Remote Sensing}, 12(24):4184, 2020.

\bibitem{yang2020basic}
Yuanxi Yang, Yue Mao, and Bijiao Sun.
\newblock {Basic Performance and Future Developments of BeiDou Global Navigation Satellite System}.
\newblock {\em Satellite Navigation}, 1(1):1, 2020.

\bibitem{zheng2023simultaneous}
Shuran Zheng, Jinling Wang, Chris Rizos, Weidong Ding, and Ahmed El-Mowafy.
\newblock {Simultaneous Localization and Mapping (SLAM) for Autonomous Driving: Concept and Analysis}.
\newblock {\em Remote Sensing}, 15(4):1156, 2023.

\bibitem{franch2020spatial}
Ivan Franch-Pardo, Brian~M Napoletano, Fernando Rosete-Verges, and Lawal Billa.
\newblock {Spatial Analysis and GIS in the Study of COVID-19. A Review}.
\newblock {\em Sci. Total Environ.}, 739:140033, 2020.

\bibitem{zhao2023seeing}
Mingmin Zhao, Peder Olsen, and Ranveer Chandra.
\newblock {Seeing Through Clouds in Satellite Images}.
\newblock {\em {IEEE} Trans. Geosci. Remote Sens.}, 61:1--16, 2023.

\bibitem{rolf2021generalizable}
Esther Rolf, Jonathan Proctor, Tamma Carleton, Ian Bolliger, Vaishaal Shankar, Miyabi Ishihara, Benjamin Recht, and Solomon Hsiang.
\newblock {A Generalizable and Accessible Approach to Machine Learning with Global Satellite Imagery}.
\newblock {\em Nat. Commun.}, 12(1):4392, 2021.

\bibitem{li2023large}
Yuxuan Li, Qibin Hou, Zhaohui Zheng, Ming-Ming Cheng, Jian Yang, and Xiang Li.
\newblock {Large Selective Kernel Network for Remote Sensing Object Detection}.
\newblock In {\em Proc. of the 18th IEEE/CVF ICCV}, pages 16794--16805, 2023.

\bibitem{lin2024efficient}
Zheng Lin, Guangyu Zhu, Yiqin Deng, Xianhao Chen, Yue Gao, Kaibin Huang, and Yuguang Fang.
\newblock {Efficient Parallel Split Learning over Resource-Constrained Wireless Edge Networks}.
\newblock {\em {IEEE} Trans. Mobile Comput.}, 23(10):9224--9239, 2024.

\bibitem{fang2024ic3m}
Zihan Fang, Zheng Lin, Senkang Hu, Hangcheng Cao, Yiqin Deng, Xianhao Chen, and Yuguang Fang.
\newblock {IC3M: In-Car Multimodal Multi-Object Monitoring for Abnormal Status of Both Driver and Passengers}.
\newblock {\em arXiv preprint arXiv:2410.02592}, 2024.

\bibitem{chen2021rf}
Zhe Chen, Chao Cai, Tianyue Zheng, Jun Luo, Jie Xiong, and Xin Wang.
\newblock {RF-Based Human Activity Recognition Using Signal Adapted Convolutional Neural Network}.
\newblock {\em {IEEE} Trans. Mobile Comput.}, 22(1):487--499, 2021.

\bibitem{peng2024sums}
Jinbo Peng, Zhe Chen, Zheng Lin, Haoxuan Yuan, Zihan Fang, Lingzhong Bao, Zihang Song, Ying Li, Jing Ren, and Yue Gao.
\newblock {SUMS: Sniffing Unknown Multiband Signals under Low Sampling Rates}.
\newblock {\em {IEEE} Trans. Mobile Comput.}, 2024.

\bibitem{yuan2025constructing}
Haoxuan Yuan, Zhe Chen, Zheng Lin, Jinbo Peng, Yuhang Zhong, Xuanjie Hu, Songyan Xue, Wei Li, and Yue Gao.
\newblock {Constructing 4D Radio Map in LEO Satellite Networks with Limited Samples}.
\newblock {\em {IEEE} INFOCOM}, 2025.

\bibitem{hu2024accelerating}
Mingda Hu, Jingjing Zhang, Xiong Wang, Shengyun Liu, and Zheng Lin.
\newblock {Accelerating Federated Learning with Model Segmentation for Edge Networks}.
\newblock {\em {IEEE} Trans. Green Commun. Netw.}, 2024.

\bibitem{tang2024merit}
Yongyang Tang, Zhe Chen, Ang Li, Tianyue Zheng, Zheng Lin, Jia Xu, Pin Lv, Zhe Sun, and Yue Gao.
\newblock {MERIT: Multimodal Wearable Vital Sign Waveform Monitoring}.
\newblock {\em arXiv preprint arXiv:2410.00392}, 2024.

\bibitem{sun2021convolutional}
Huiming Sun, Yuewei Lin, Qin Zou, Shaoyue Song, Jianwu Fang, and Hongkai Yu.
\newblock {Convolutional Neural Networks Based Remote Sensing Scene Classification under Clear and Cloudy Environments}.
\newblock In {\em Proc. of the 34th IEEE/CVF CVPR}, pages 713--720, 2021.

\bibitem{yuan2023graph}
Haoxuan Yuan, Zhe Chen, Zheng Lin, Jinbo Peng, Zihan Fang, Yuhang Zhong, Zihang Song, Xiong Wang, and Yue Gao.
\newblock {Graph Learning for Multi-Satellite Based Spectrum Sensing}.
\newblock In {\em {Proc. IEEE Int. Conf. Commun. Technol. (ICCT)}}, pages 1112--1116, 2023.

\bibitem{lin2025hasfl}
Zheng Lin, Zhe Chen, Xianhao Chen, Wei Ni, and Yue Gao.
\newblock {HASFL: Heterogeneity-Aware Split Federated Learning over Edge Computing Systems}.
\newblock {\em arXiv preprint arXiv:2506.08426}, 2025.

\bibitem{wu2024netllm}
Duo Wu, Xianda Wang, Yaqi Qiao, Zhi Wang, Junchen Jiang, Shuguang Cui, and Fangxin Wang.
\newblock {Netllm: Adapting Large Language Models for Networking}.
\newblock In {\em Proc. of the 34th ACM SIGCOMM}, pages 661--678, 2024.

\bibitem{zhang2025lcfed}
Yuxin Zhang, Haoyu Chen, Zheng Lin, Zhe Chen, and Jin Zhao.
\newblock {LCFed: An Efficient Clustered Federated Learning Framework for Heterogeneous Data}.
\newblock {\em arXiv preprint arXiv:2501.01850}, 2025.

\bibitem{chen2024gradient}
Haoyu Chen, Yuxin Zhang, Jin Zhao, Xin Wang, and Yuedong Xu.
\newblock Gradient free personalized federated learning.
\newblock In {\em Proceedings of the 53rd International Conference on Parallel Processing}, pages 971--980, 2024.

\bibitem{lin2024adaptsfl}
Zheng Lin, Guanqiao Qu, Wei Wei, Xianhao Chen, and Kin~K Leung.
\newblock {Adaptsfl: Adaptive Split Federated Learning in Resource-Constrained Edge Networks}.
\newblock {\em {IEEE} Trans. Netw.}, 2024.

\bibitem{kaplan2020scaling}
Jared Kaplan, Sam McCandlish, Tom Henighan, Tom~B Brown, Benjamin Chess, Rewon Child, Scott Gray, Alec Radford, Jeffrey Wu, and Dario Amodei.
\newblock {Scaling Laws for Neural Language Models}.
\newblock {\em arXiv preprint arXiv:2001.08361}, 2020.

\bibitem{10490262}
Danfeng Hong, Bing Zhang, Xuyang Li, Yuxuan Li, Chenyu Li, Jing Yao, Naoto Yokoya, Hao Li, Pedram Ghamisi, Xiuping Jia, Antonio Plaza, Paolo Gamba, Jon~Atli Benediktsson, and Jocelyn Chanussot.
\newblock {SpectralGPT: Spectral Remote Sensing Foundation Model}.
\newblock {\em {IEEE} Trans. Pattern Anal. Mach. Intell.}, 46(8):5227--5244, 2024.

\bibitem{lin2025hsplitlora}
Zheng Lin, Yuxin Zhang, Zhe Chen, Zihan Fang, Xianhao Chen, Praneeth Vepakomma, Wei Ni, Jun Luo, and Yue Gao.
\newblock {HSplitLoRA: A Heterogeneous Split Parameter-Efficient Fine-Tuning Framework for Large Language Models}.
\newblock {\em arXiv preprint arXiv:2505.02795}, 2025.

\bibitem{liu2024remoteclip}
Fan Liu, Delong Chen, Zhangqingyun Guan, Xiaocong Zhou, Jiale Zhu, Qiaolin Ye, Liyong Fu, and Jun Zhou.
\newblock {RemoteCLIP: A Vision Language Foundation Model for Remote Sensing}.
\newblock {\em {IEEE} Trans. Geosci. Remote Sens.}, 2024.

\bibitem{zhang2024earthgpt}
Wei Zhang, Miaoxin Cai, Tong Zhang, Yin Zhuang, and Xuerui Mao.
\newblock {EarthGPT: A Universal Multimodal Large Language Model for Multisensor Image Comprehension in Remote Sensing Domain}.
\newblock {\em {IEEE} Trans. Geosci. Remote Sens.}, 62:1--20, 2024.

\bibitem{zhan2024skyeyegpt}
Yang Zhan, Zhitong Xiong, and Yuan Yuan.
\newblock {SkyEyeGPT: Unifying Remote Sensing Vision-Language Tasks via Instruction Tuning with Large Language Model}.
\newblock {\em arXiv preprint arXiv:2401.09712}, 2024.

\bibitem{zhang2024vision}
Jingyi Zhang, Jiaxing Huang, Sheng Jin, and Shijian Lu.
\newblock {Vision-Language Models for Vision Tasks: A Survey}.
\newblock {\em {IEEE} Trans. Pattern Anal. Mach. Intell.}, 46(8):5625--5644, 2024.

\bibitem{zhang2024pip}
Yudong Zhang, Ruobing Xie, Jiansheng Chen, Xingwu Sun, and Yu~Wang.
\newblock {PIP: Detecting Adversarial Examples in Large Vision-Language Models via Attention Patterns of Irrelevant Probe Questions}.
\newblock In {\em Proc. of the 32nd ACM MM}, pages 11175--11183, 2024.

\bibitem{wang2024break}
Yubo Wang, Chaohu Liu, Yanqiu Qu, Haoyu Cao, Deqiang Jiang, and Linli Xu.
\newblock {Break the Visual Perception: Adversarial Attacks Targeting Encoded Visual Tokens of Large Vision-Language Models}.
\newblock In {\em Proc. of the 32nd ACM MM}, pages 1072--1081, 2024.

\bibitem{fang2024automated}
Zihan Fang, Zheng Lin, Zhe Chen, Xianhao Chen, Yue Gao, and Yuguang Fang.
\newblock {Automated Federated Pipeline for Parameter-Efficient Fine-Tuning of Large Language Models}.
\newblock {\em arXiv preprint arXiv:2404.06448}, 2024.

\bibitem{lin2024splitlora}
Zheng Lin, Xuanjie Hu, Yuxin Zhang, Zhe Chen, Zihan Fang, Xianhao Chen, Ang Li, Praneeth Vepakomma, and Yue Gao.
\newblock {SplitLoRA: A Split Parameter-Efficient Fine-Tuning Framework for Large Language Models}.
\newblock {\em arXiv preprint arXiv:2407.00952}, 2024.

\bibitem{ramesh2021zero}
Aditya Ramesh, Mikhail Pavlov, Gabriel Goh, Scott Gray, Chelsea Voss, Alec Radford, Mark Chen, and Ilya Sutskever.
\newblock {Zero-Shot Text-to-Image Generation}.
\newblock In {\em Proc. of the 38th ICML}, pages 8821--8831, 2021.

\bibitem{kuckreja2024geochat}
Kartik Kuckreja, Muhammad~Sohail Danish, Muzammal Naseer, Abhijit Das, Salman Khan, and Fahad~Shahbaz Khan.
\newblock {GeoChat: Grounded Large Vision-Language Model for Remote Sensing}.
\newblock In {\em Proc. of the 37th IEEE/CVF CVPR}, pages 27831--27840, 2024.

\bibitem{denby2020orbital}
Bradley Denby and Brandon Lucia.
\newblock {Orbital Edge Computing: Nanosatellite Constellations as a New Class of Computer System}.
\newblock In {\em Proc. of the 25th ACM ASPLOS}, pages 939--954, 2020.

\bibitem{raspberrypi}
``{{Small Satellites and Big Antennas}}", 2023.
\newblock Available: https://www.raspberrypi.com/news/small-satellites-and-big-antennas.

\bibitem{wang2024qwen2}
Peng Wang, Shuai Bai, Sinan Tan, Shijie Wang, Zhihao Fan, Jinze Bai, Keqin Chen, Xuejing Liu, Jialin Wang, Wenbin Ge, et~al.
\newblock {Qwen2-VL: Enhancing Vision-Language Model’s Perception of the World at Any Resolution}.
\newblock {\em arXiv preprint arXiv:2409.12191}, 2024.

\bibitem{zhang2025s}
Yuxin Zhang, Zhe Chen, Xuanjie Hu, Jin Zhao, and Yue Gao.
\newblock S-leon: An efficient split learning framework over heterogeneous leo satellite networks.
\newblock {\em Authorea Preprints}, 2025.

\bibitem{vasisht2021l2d2}
Deepak Vasisht, Jayanth Shenoy, and Ranveer Chandra.
\newblock {L2D2: Low Latency Distributed Downlink for LEO Satellites}.
\newblock In {\em Proc. of the 35th ACM SIGCOMM}, pages 151--164, 2021.

\bibitem{xia2018dota}
Gui-Song Xia, Xiang Bai, Jian Ding, Zhen Zhu, Serge Belongie, Jiebo Luo, Mihai Datcu, Marcello Pelillo, and Liangpei Zhang.
\newblock {DOTA: A Large-Scale Dataset for Object Detection in Aerial Images}.
\newblock In {\em Proc. of the 31st IEEE CVPR}, pages 3974--3983, 2018.

\bibitem{murphy2024deploying}
James Murphy, Maria Buckley, Leonie Buckley, Adam Taylor, Jake O'brien, and Brian Mac~Namee.
\newblock {Deploying Machine Learning Anomaly Detection Models to Flight Ready AI Boards}.
\newblock In {\em Proc. of the 41st IEEE/CVF CVPR}, pages 6828--6836, 2024.

\bibitem{manning2018machine}
Jacob Manning, David Langerman, Barath Ramesh, Evan Gretok, Christopher Wilson, Alan George, James MacKinnon, and Gary Crum.
\newblock {Machine-Learning Space Applications on SmallSat Platforms with TensorFlow}.
\newblock 2018.

\bibitem{george2018onboard}
Alan~D George and Christopher~M Wilson.
\newblock {Onboard Processing With Hybrid and Reconfigurable Computing on Small Satellites}.
\newblock {\em Proc. {IEEE}}, 106(3):458--470, 2018.

\bibitem{10769058}
Peng Xu, Wenqi Shao, Kaipeng Zhang, Peng Gao, Shuo Liu, Meng Lei, Fanqing Meng, Siyuan Huang, Yu~Qiao, and Ping Luo.
\newblock {LVLM-EHub: A Comprehensive Evaluation Benchmark for Large Vision-Language Models}.
\newblock {\em {IEEE} Trans. Pattern Anal. Mach. Intell.}, pages 1--18, 2024.

\bibitem{lobry2020rsvqa}
Sylvain Lobry, Diego Marcos, Jesse Murray, and Devis Tuia.
\newblock {RSVQA: Visual Question Answering for Remote Sensing Data}.
\newblock {\em {IEEE} Trans. Geosci. Remote Sens.}, 58(12):8555--8566, 2020.

\bibitem{cheng2017remote}
Gong Cheng, Junwei Han, and Xiaoqiang Lu.
\newblock {Remote Sensing Image Scene Classification: Benchmark and State of the Art}.
\newblock {\em Proc. {IEEE}}, 105(10):1865--1883, 2017.

\bibitem{jetson}
``{{Planet Labs PBC Announces Real-Time Insights Technology Using NVIDIA Jetson Platform}}", 2024.
\newblock Available: https://www.businesswire.com/news/home/20240610385569/en/Planet-Labs-PBC-Announces-Real-Time-Insights-Technology-Using-NVIDIA-Jetson-Platform.

\bibitem{10.1145/3570361.3592521}
Bill Tao, Maleeha Masood, Indranil Gupta, and Deepak Vasisht.
\newblock {Transmitting, Fast and Slow: Scheduling Satellite Traffic Through Space and Time}.
\newblock In {\em Proc. of the 29th ACM MobiCom}, pages 1--15, 2023.

\bibitem{starlink_gp}
``{{NORAD GP Element Sets}}", 2024.
\newblock Available: https://celestrak.org/NORAD/elements/.

\bibitem{zhong2024urbancross}
Siru Zhong, Xixuan Hao, Yibo Yan, Ying Zhang, Yangqiu Song, and Yuxuan Liang.
\newblock {UrbanCross: Enhancing Satellite Image-Text Retrieval with Cross-Domain Adaptation}.
\newblock In {\em Proc. of the 32nd ACM MM}, pages 6307--6315, 2024.

\bibitem{pfaff2015design}
Ben Pfaff, Justin Pettit, Teemu Koponen, Ethan Jackson, Andy Zhou, Jarno Rajahalme, Jesse Gross, Alex Wang, Joe Stringer, Pravin Shelar, et~al.
\newblock {The Design and Implementation of Open vSwitch}.
\newblock In {\em Proc. of th 12th NSDI}, pages 117--130, 2015.

\bibitem{beshay2015fidelity}
Joseph~D Beshay, Andrea Francini, and Ravi Prakash.
\newblock {On the Fidelity of Single-Machine Network Emulation in Linux}.
\newblock In {\em Proc. of the 23rd IEEE MASCOTS}, pages 19--22, 2015.

\bibitem{radford2021learning}
Alec Radford, Jong~Wook Kim, Chris Hallacy, Aditya Ramesh, Gabriel Goh, Sandhini Agarwal, Girish Sastry, Amanda Askell, Pamela Mishkin, Jack Clark, et~al.
\newblock {Learning Transferable Visual Models From Natural Language Supervision}.
\newblock In {\em Proc. of the 38th ICML}, pages 8748--8763, 2021.

\bibitem{wang2023tabi}
Yiding Wang, Kai Chen, Haisheng Tan, and Kun Guo.
\newblock {Tabi: An Efficient Multi-Level Inference System for Large Language Models}.
\newblock In {\em Proc. of the 18th ECCS EuroSys}, pages 233--248, 2023.

\bibitem{he2024large}
Ying He, Jingcheng Fang, F~Richard Yu, and Victor~C Leung.
\newblock {Large Language Models (LLMs) Inference Offloading and Resource Allocation in Cloud-Edge Computing: An Active Inference Approach}.
\newblock {\em {IEEE} Trans. Mob. Comput.}, 23(12):11253--11264, 2024.

\bibitem{hu2023rsgpt}
Yuan Hu, Jianlong Yuan, Congcong Wen, Xiaonan Lu, and Xiang Li.
\newblock {RSGPT: A Remote Sensing Vision Language Model and Benchmark}.
\newblock {\em arXiv preprint arXiv:2307.15266}, 2023.

\bibitem{yao2022edge}
Jiangchao Yao, Shengyu Zhang, Yang Yao, Feng Wang, Jianxin Ma, Jianwei Zhang, Yunfei Chu, Luo Ji, Kunyang Jia, Tao Shen, et~al.
\newblock {Edge-Cloud Polarization and Collaboration: A Comprehensive Survey for AI}.
\newblock {\em {IEEE} Trans. Knowl. Data Eng.}, 35(7):6866--6886, 2022.

\bibitem{zhang2024fedac}
Yuxin Zhang, Haoyu Chen, Zheng Lin, Zhe Chen, and Jin Zhao.
\newblock {FedAC: An Adaptive Clustered Federated Learning Framework for Heterogeneous Data}.
\newblock {\em arXiv preprint arXiv:2403.16460}, 2024.

\bibitem{borzunov2022petals}
Alexander Borzunov, Dmitry Baranchuk, Tim Dettmers, Max Ryabinin, Younes Belkada, Artem Chumachenko, Pavel Samygin, and Colin Raffel.
\newblock {Petals: Collaborative Inference and Fine-tuning of Large Models}.
\newblock {\em arXiv preprint arXiv:2209.01188}, 2022.

\end{thebibliography}

\end{document}